\documentclass[preprint,pra,aps,floatfix,showpacs]{revtex4}

\usepackage{graphics}
\usepackage{bm}

\def\PRA{{Phys.~Rev.~A} }

\def\JPB{{J.~Phys.~B} }
\def\PRL{{Phys.~Rev.~Lett.} }
\def\RMP{{Rev.~Mod.~Phys.} }
\def\JCP{{J.~Chem.~Phys.} }

\newcommand{\myscalebox}[1]{\scalebox{0.35}[0.35]{#1}}

\newcommand{\myscaleboxc}[1]{\scalebox{0.4}[0.4]{#1}}
\newcommand{\myscaleboxd}[1]{\scalebox{0.75}[0.75]{#1}}

\newcommand{\be}{\begin{equation}}
\newcommand{\bea}{\begin{eqnarray}}
\newcommand{\eea}{\end{eqnarray}}
\newcommand{\ee}{\end{equation}}

\newcommand{\bra}{{\langle}}
\newcommand{\ket}{{\rangle}}

\begin{document}

\title{Quantitative Rescattering Theory for high-order harmonic generation
from molecules}

\author{Anh-Thu Le,$^1$
R.~R. Lucchese,$^2$ S. Tonzani,$^3$ T. Morishita,$^4$ and C.~D.
Lin$^1$}

\affiliation{$^1$Department of Physics, Cardwell Hall, Kansas
State University, Manhattan, KS 66506, USA\\
$^2$Department of Chemistry, Texas A\&M University, College Station,
Texas 77843-3255, USA\\
$^3$Nature Publishing Group 4 Crinan Street, London N1 9XW, UK\\
$^4$Department of Applied Physics and Chemistry, University of
Electro-Communications, 1-5-1 Chofu-ga-oka, Chofu-shi, Tokyo,
182-8585, Japan and PRESTO, JST Agency, Kawaguchi, Saitama 332-0012,
Japan}

\date{\today}

\begin{abstract}
The Quantitative Rescattering Theory (QRS) for high-order harmonic
generation (HHG) by intense laser pulses is presented. According to
the QRS, HHG spectra can be expressed as a product of a returning
electron wave packet and the photo-recombination differential cross
section of the {\em laser-free} continuum electron back to the
initial bound state. We show that the shape of the returning
electron wave packet is determined mostly by the laser only.  The
returning electron wave packets can be obtained from the
strong-field approximation or from the solution of the
time-dependent Schr\"odinger equation (TDSE) for a reference atom.
The validity of the QRS is carefully examined by checking against
accurate results for both harmonic magnitude and phase from the
solution of the TDSE for atomic targets within the single active
electron approximation. Combining with accurate transition dipoles
obtained from state-of-the-art molecular photoionization
calculations, we further show that available experimental
measurements for HHG from partially aligned molecules can be
explained by the QRS. Our results show that quantitative description
of the HHG from aligned molecules has become possible. Since
infrared lasers of pulse durations of a few femtoseconds are easily
available in the laboratory, they may be used for dynamic imaging of
a transient molecule with femtosecond temporal resolutions.
\end{abstract}

\pacs{42.65.Ky, 33.80.Rv}
\maketitle

\section{Introduction}

High-order harmonic generation (HHG) has been studied extensively
since 1990's, both experimentally and theoretically. Initial
interest in HHG was related to the generation of coherent soft X-ray
beams \cite{krausz97,kapteyn97}, which are currently being used for
many applications in ultrafast science experiments
\cite{krausz09,thomann08}. In the past decade, HHG has also been
used for the production of single attosecond pulses
\cite{drescher,sekikawa,sansone} and attosecond pulse trains
\cite{apt}, thus opening up new opportunities for attosecond
time-resolved spectroscopy.

While HHG has been well studied for atoms, much less has been done
for molecules. Initial interest in HHG from molecules was due to the
fact that it offers promising prospects for increasing the low
conversion efficiency for harmonic generation. The presence of the
additional degrees of freedom such as the alignment and vibration
opens possibilities of controlling the phase of the nonlinear
polarization of the medium and of meeting the phase-matching
condition. Thanks to the recent advancements in molecular alignment
and orientation techniques \cite{seideman}, investigation of the
dependence of HHG on molecular alignment reveals the distinctive
features of molecular HHG \cite{itatani05} as well as their
structure \cite{itatani-nature}. In particular, the existence of
distinctive minima in the HHG spectra from H$_2^+$ have been
theoretically predicted by Lein {\it et al.} \cite{lein02,lein04}.
The experimental measurements within the pump-probe scheme for
CO$_2$ indeed showed the signature of these minima for partially
aligned CO$_2$ \cite{kanai05,vozzi05}. More recently, newer
experiments by JILA, Saclay, and Riken groups
\cite{jila07,jila08,boutu,riken08} began to focus on the phase of
the harmonics, using mixed gases and interferometry techniques.
These studies have revealed that near the harmonic yield minimum the
phase undergoes a big change.  In view of these experimental
developments, a quantitative theory for HHG from molecular targets
is timely.

HHG can be understood using the three-step model
\cite{corkum,kulander}. First the electron is released into
continuum by tunnel ionization; second, it is accelerated by the
oscillating electric field of the laser and later driven back to the
target ion; and third, the electron recombines with the ion to emit
a high energy photon. A semiclassical formulation of this three-step
model based on the strong field approximation (SFA) is given by
Lewenstein {\it et al} \cite{lewenstein}. In SFA, the liberated
continuum electron experiences the full effect from the laser field,
but not from the ion. In spite of this limitation, the model has
been widely used for understanding the HHG by atoms and molecules.
Since the continuum electron needs to come back to revisit the
parent ion in order to emit radiation, the neglect of electron-ion
interaction is clearly questionable. Thus, over the years efforts
have been made to improve upon the SFA model, by including Coulomb
distortion \cite{ivanov96,kaminski96,smirnova08}. These
improvements, when applied to simple systems, however, still do not
lead to satisfactory agreement with accurate calculations, based on
numerical solution of the time-dependent Schr\"odinger equation
(TDSE).

With moderate effort, direct numerical solution of the TDSE for
atomic targets can be carried out, at least within the single active
electron (SAE) approximation. For molecules, accurate numerical
solution of the TDSE is much more computational demanding and has
not been carried out except for the simplest molecules such as
H$_2^+$ \cite{lein02,kamta05,telnov07}. Thus most of the existing
calculations for HHG from molecules were performed using the SFA
model
\cite{zhou05,zhou05b,atle06,atle07,marangos06,seideman07,milosevic09,faisal09}.
Additionally, in order to compare single molecule calculations with
experimental measurements, macroscopic propagation of the emitted
radiation field in a gas jet or chamber needs to be carried out,
with input from calculations involving hundreds of laser intensities
to account for intensity variation near the laser focus. For this
purpose the TDSE method is clearly too time-consuming even for
atomic targets.

In view of the inaccuracies of the SFA and the practical
inefficiency of the TDSE method, we have proposed the quantitative
rescattering theory or the QRS (also called the scattering-wave SFA
or SW-SFA in our earlier papers \cite{atle08,H2+}) as a simple and
practical method for obtaining accurate HHG spectra. The main idea
of the QRS is to employ exact transition dipoles with scattering
wave instead of the commonly used plane-waves for the recombination
step in a HHG process. Within the QRS, induced dipole moment by the
laser can be represented as a product of the returning electron wave
packet and the {\em complex} recombination transition dipole between
free electrons with the atomic or molecular ion \cite{atle08,toru}.
The shape of the wave packet has been shown to be largely
independent of the target. Therefore the returning wave  packet can
be calculated from the standard SFA model or by solving the TDSE
from a reference atom with similar ionization potential as of the
molecule under consideration. The QRS approach has been shown to be
much more accurate than the standard SFA model for rare-gas atoms
\cite{atle08} and a prototypical molecular system H$_2^+$
\cite{H2+}. We note that the QRS has also been successfully used to
explain high-energy electron spectra from various atomic systems
\cite{chen-jpb09,chen09,sam-prl09,sam-jpb09,chen07}, and
non-sequential double ionization \cite{sam-nsdi} of atoms.

In order to use the QRS, we need to know the {\em complex}
transition dipoles from aligned molecules. In other words, we need
both the differential photoionization (or photo-recombination) cross
section and the phase of the transition dipole for a fixed-in-space
molecule. Calculations for these quantities can be carried out using
the available molecular photoionization methods, which have been
developed over the last few decades. In this paper we employ the
iterative Schwinger variational method developed by Lucchese and
collaborators \cite{lucchese82,lucchese95}.

The goal of this paper is to give a detailed description of all the
ingredients of the QRS and demonstrate its validity. As examples of its
application we consider HHG from aligned O$_2$ and CO$_2$ and show
that the QRS is capable of explaining recent experiments. The rest
of this paper is arranged as follows. In Sec.~II we summarize all
the theoretical tools needed for calculating HHG spectra from
partially aligned molecules. We also explain how photoionization
(photo-recombination) differential cross sections and phases of the
transitions dipole are calculated for both atomic and molecular
targets. The QRS model is formulated and its validity is carefully
examined in Sec.~III for atomic targets where results from the TDSE
serve as ``experimental'' data. In Sec.~IV we illustrate the
application of the QRS method by considering the two examples of O$_2$ and CO$_2$ and
compare the QRS results with available experiments. Sec.~V discusses
the predictions of the QRS for the minima in HHG spectra in relation
to the simple two-center interference model. We also extend the
QRS calculation for O$_2$ to higher photon energies by using 1600-nm
lasers, to explore the interference effect in HHG spectra from
O$_2$.  Finally we finish our paper with a summary and outlook.

We note that the macroscopic propagation effect has not been treated
in this paper \cite{jin09}. In the literature, theoretical
treatments of such effects have been limited to atomic targets, and
mostly starting with the SFA to calculate  single atom response to
the lasers. In this respect, the QRS offers an attractive
alternative as the starting point since its calculation of single
atom response is nearly as fast as the SFA, but with an accuracy
much closer to the TDSE. Atomic units are used throughout this paper
unless otherwise stated.

\section{Theoretical background}
The theory part is separated into six subsections. We will describe
all the ingredients used in the QRS. First we consider the TDSE and
the SFA, as the methods to calculate HHG spectra and to extract
returning electron wave packets. We will then present theoretical
methods for calculation of photoionization (photo-recombination)
cross sections for both atomic and linear molecular targets. We will
also briefly describe intense laser ionization, as ionization rates
are used in the QRS to account for depletion effect and overall
normalization of the wave packets. Lastly, we explain how the
partial alignment of molecules by an aligning laser is treated.

\subsection{Method of solving the time-dependent Schr\"{o}dinger equation}

The method for solving the time-dependent Schr\"{o}dinger equation
(TDSE) for an atom in an intense laser pulse has been described in
our previous works \cite{chen06,toru07,chen09}. Here we present only
the essential steps of the calculations and the modifications needed
to treat the HHG problem. At present accurate numerical solution of
the TDSE for molecules is still a formidable computational task and
has been carried out mostly for simple molecular systems such as
H$_2^+$ \cite{lein02,kamta05,telnov07}.

We treat the target atom in the single active electron model. The
Hamiltonian for such an atom in the presence of a linearly polarized
laser pulse can be written as
\begin{eqnarray}
\label{tdse2} H=H_0+H_i(t)=-\frac{1}{2}\nabla^2 +V(r)+H_i(t).
\end{eqnarray}
The atomic model potential $V(r)$ is parameterized in the form
\begin{eqnarray}
\label{model-pot}V(r)=-\frac{1+a_{1}e^{-a_{2}r}+a_{3}re^{-a_{4}r}+a_{5}e^{-a_{6}r}}{r},
\end{eqnarray}
which can also be written as the sum of a short-range potential and an attractive Coulomb potential
\begin{eqnarray}
\label{short-pot}V(r)=-\frac{1}{r} + V_s(r),
\end{eqnarray}
The parameters in Eq.~(\ref{model-pot}) are obtained by fitting the
calculated binding energies from this potential to the experimental
binding energies of the ground state and the first few excited
states of the target atom. The parameters for the targets used in
this paper can be found in \cite{tong-jpb05}. One can also use a
scaled hydrogen as a reference atom, as shown in
Ref.~\cite{atle08,H2+}. Here the effective nuclear charge is
rescaled such that the ionization potential of the rescaled H(1s)
matches the $I_p$ of the atom or molecule under consideration.

The electron-field interaction $H_i(t)$, in length gauge, is given by
\begin{eqnarray}
\label{tdse3}H_i(t)={\bm r}\cdot{\bm E}(t).
\end{eqnarray}
For a linearly polarized laser pulse (along the $z$ axis) with
carrier frequency $\omega$ and carrier-envelope-phase (CEP),
$\varphi$, the electric field is taken to have the form
\begin{eqnarray}
{\bm E}(t)=\hat{z}E_{0}\cos^{2}\left(\frac{\pi
t}{\tau}\right)\cos(\omega t+\varphi)
\end{eqnarray}
for the time interval ($-\tau$/2, $\tau$/2) and zero elsewhere. The
pulse duration, defined as the full width at half maximum (FWHM) of
the intensity, is given by $\Gamma=\tau/2.75$.

The time evolution of the electronic wavefunction
$\Psi(\textbf{r},t)$, which satisfies the TDSE,
\begin{eqnarray}
\label{tdse1}i\frac{\partial}{\partial t}\Psi({\bm r},t)=H\Psi({\bm
r},t)
\end{eqnarray}
is solved by expanding in terms of eigenfunctions,
$R_{nl}(r)Y_{lm}(\mathbf{\hat{r}})$, of $H_0$, within the box of
$r\in[0,r_{\text{max}}]$
\begin{eqnarray}
\label{expansion}\Psi({\bm
r},t)=\sum_{nl}c_{nl}(t)R_{nl}(r)Y_{lm}(\mathbf{\hat{r}})
\end{eqnarray}
where radial functions $R_{nl}(r)$ are expanded in a discrete
variable representation (DVR) basis set \cite{DVR} associated with
Legendre polynomials, while $c_{nl}$ are calculated using the
split-operator method ~\cite{tong97}
\begin{eqnarray}
\label{split-operator}c_{nl}(t+\Delta t)&\simeq&
\sum_{n'l'}\{e^{-iH_0 \Delta
t/2}e^{-iH_i(t+\Delta t/2)\Delta t} \nonumber \\
&&\times e^{-iH_0 \Delta t/2}\}_{nl,n'l'} c_{n'l'}(t).
\end{eqnarray}
For rare gas atoms, which have p-wave ground state wavefunctions,
only $m=0$ is taken into account since for linearly polarized laser
pulses, contribution to the ionization probability from $m=\pm1$ is
much smaller in comparison to the $m=0$ component.

Once the time-dependent wavefunction is determined, one can
calculate induced dipole in either the length or acceleration forms

\begin{equation}
D_L(t)=\langle\Psi({\bm r},t)|z|\Psi({\bm r},t)\rangle
\end{equation}
\begin{equation}
D_A(t)=\langle\Psi({\bm r},t)|\frac{\partial V(r)}{\partial z}
|\Psi({\bm r},t)\rangle
\end{equation}
For low intensities, when the ionization is insignificant, the two
forms agree very well. For higher intensities the above length form
should be corrected by a boundary term, to account for the
non-negligible amount of electron escape to infinity, see Burnett
{\it et al} \cite{burnett}. Therefore, we found it more convenient
to use the acceleration form. To avoid artificial reflection due to
a finite box-size, we use an absorber of the form
$\cos^{1/4}[\pi(r-r_{cut})/2(r_{max}-r_{cut})]$ for $r\geq r_{cut}$
\cite{tong97}, to filter out the wave packet reaching the boundary.
We typically use $r_{max}=200$ to 400, $r_{cut}\approx r_{max}-100$,
with up to about 800 DVR points and 80 partial-waves. We have
checked that the results are quite insensitive with respect to the
absorber parameter $r_{cut}$.

\subsection{The strong field approximation (SFA)}

The strong-field approximation (SFA) has been widely used for
theoretical simulations of HHG from atoms \cite{lewenstein} and
molecules
\cite{zhou05,zhou05b,atle06,atle07,marangos06,seideman07,milosevic09,faisal09}.
This model is known to give qualitatively good results, especially
for harmonics near the cutoff. However, in the lower plateau region
the SFA model is not accurate \cite{atle08}. Nevertheless, since the
propagation of electrons after tunnel ionization is dominated by the
laser field, one can use the SFA to extract quite accurate returning
electron wave packet which can be used in the QRS theory. Here we
briefly describe the SFA, extended for molecular targets
\cite{zhou05}.

Without loss of generality, we assume that the molecules are aligned
along the $x$-axis, in a laser field $E(t)$, linearly polarized on the
$x$-$y$ plane with an angle $\theta$ with respect to the molecular
axis. The parallel component of the induced dipole moment can be
written in the form
\begin{eqnarray}
D_{\parallel}(t)& = &
i\int_0^{\infty}d\tau\left(\frac{\pi}{\epsilon+i\tau/2}
\right)^{3/2}[\cos\theta d^*_x(t)+\sin\theta d^*_y(t)]\nonumber\\
 & &
\times[\cos\theta d_x(t-\tau)+\sin\theta d_y(t-\tau)]E(t-\tau)\nonumber\\
& & \times\exp[-iS_{st}(t,\tau)]a^*(t)a(t-\tau)+c.c.
\label{lewenstein}
\end{eqnarray}
where ${\bm d}(t)\equiv{\bm d} [{\bm p}_{st}(t,\tau)+{\bm A}(t)]$,
${\bm d}(t-\tau)\equiv{\bm d} [{\bm p}_{st}(t,\tau)+{\bm
A}(t-\tau)]$ are the transition dipole moments between the ground
state and the continuum state, and ${\bm p}_{st}(t,\tau) =
-\int_{t-\tau}^{t} \bm{A}(t') dt'/\tau$ is the canonical momentum at
the stationary points, with ${\bm A}$ the vector potential. The
perpendicular component $D_{\perp}(t)$ is given by a similar formula
with $[\cos\theta d^*_x(t)+\sin\theta d^*_y(t)]$ replaced by
$[\sin\theta d^*_x(t)-\cos\theta d^*_y(t)]$ in
Eq.~(\ref{lewenstein}). The action at the stationary points for the
electron propagating in the laser field is
\begin{equation}
S_{st}(t,\tau) = \int_{t-\tau}^{t} \left( \frac{
[\bm{p}_{st}(t,\tau)+\bm{A}(t')]^2}{2}+I_p\right) dt',
\label{action}
\end{equation}
where $I_p$ is the ionization potential of the molecule. In
Eq.~(\ref{lewenstein}), $a(t)$ is introduced to account for the
ground state depletion.

The HHG power spectra are obtained from Fourier components of the
induced dipole moment $D(t)$ as given by
\begin{equation}
P(\omega)\propto
|a(\omega)|^2=\left|\int\frac{d^2D(t)}{dt^2}e^{i\omega t} dt
\right|^2 \approx \omega^4|D(\omega)|^2.
\end{equation}

In our calculations we use ground state electronic wavefunctions
obtained from the general quantum chemistry codes such as Gamess
\cite{gamess} or Gaussian \cite{gaussian}. Within the single active
electron (SAE) approximation, we take the highest occupied molecular
orbital (HOMO) for the ``ground state''. In the SFA the transition
dipole ${\bm d}({\bm k})$ is given as $\langle{\bm k}|{\bm
r}|\Psi_0\rangle$ with the continuum state approximated by a
plane-wave $|{\bm k}\rangle$. In order to account for the depletion
of the ground state, we approximate the ground state amplitude by
$a(t)=\exp[-\int_{-\infty}^t W(t')/2 dt']$, with the ionization rate
$W(t')$ obtained from the MO-ADK theory or the MO-SFA [see,
Sec.~II(E)].

\subsection{Calculation of transition dipoles for atomic targets}

Photo-recombination process is responsible for the last step in the
three-step model for HHG. The QRS model goes beyond the standard
plane-wave approximation (PWA) and employs exact scattering wave to
calculate the transition dipoles. Since photo-recombination is the
time-reversed process of photoionization, in this and the next
subsection we will analyze the basic formulations and methods for
calculating both processes in atoms and linear molecules.

The photoionization cross section for transition from an initial
bound state $\Psi_i$ to the final continuum state $\Psi_{{\bm k}}^-$
due to a linearly polarized light field is proportional to the
modulus square of the transition dipole (in the length form)
\begin{equation}
d_{{\bm k},{\bm n}}(\omega)=\langle\Psi_i|{\bm r}\cdot{\bm
n}|\Psi_{{\bm k}}^-\rangle. \label{dipole}
\end{equation}
Here ${\bm n}$ is the direction of the light polarization and ${\bm
k}$ is the momentum of the ejected photoelectron. The
photoionization differential cross section (DCS) can be expressed in
the general form as \cite{starace}
\begin{eqnarray}
\frac{d^2\sigma^I}{d\Omega_{\bm k}d\Omega_{\bm n}}=
\frac{4\pi^2\omega k}{c} \left|\bra\Psi_i|{\bm r}\cdot{\bm
n}|\Psi_{\bm k}^{-}\ket \right|^2, \label{photo-DCS}
\end{eqnarray}
where $k^2/2+I_p=\omega$ with $I_p$ being the ionization potential,
$\omega$ the photon energy, and $c$ the speed of light.

To be consistent with the treatment of the TDSE for atoms in laser
fields in Sec.~II(A), we will use the model potential approach for
atomic targets. The continuum wavefunction $\Psi_{\bm k}^{-}({\bf
r})$ then satisfies the Schr\"odinger equation
\begin{equation}
\left[-\frac{\nabla^2}{2}+V(r)-\frac{k^2}{2}\right]\Psi_{{\bm
k}}^-({\bm r})=0, \label{scat-Eq}
\end{equation}
where the spherically symmetric model potential $V(r)$ for each
target is the same as in Eq.~(\ref{model-pot}).

The incoming scattering wave can be expanded in terms of partial
waves as
\begin{eqnarray}
\Psi_{\bm k}^{-}({\bm r})&=&\frac{1}{\sqrt{k}}
\sum_{l=0}^{\infty}\sum_{m=-l}^l i^l \exp[-i(\sigma_l+\delta_l)]
R_{El}(r) Y_{lm}(\Omega_{\bm r}) Y_{lm}^*(\Omega_{\bm k}).
\label{PW-expansion}
\end{eqnarray}
Here, $\delta_l$ is the $l$-th partial wave phase shift due to the
short range  potential $V_s(r)$ in Eq.~(\ref{model-pot}), and
$\sigma_l$ is the Coulomb phase shift
\begin{eqnarray}
\sigma_l&=&\arg\Gamma(l+1+i\gamma),\\
\gamma&=&-Z/k,
\end{eqnarray}
with the asymptotic nuclear charge $Z=1$. $R_{El}$ is the energy
normalized radial wavefunction such that
\begin{equation}
\int_0^{\infty} R_{E l}(r) R_{E' l}(r) r^2 dr=\delta(E-E'),
\end{equation}
and has the asymptotic form
\begin{equation}
R_{E l}(r)\rightarrow\frac{1}{r}\sqrt{\frac{2}{\pi k}} \sin(kr -
l\pi/2 -\gamma \log 2 kr+\sigma_l+\delta_l).
\end{equation}

The initial bound state can be written as
\begin{equation}
\Psi_i({\bm r})=R_{nl_i}(r)Y_{l_i m_i}(\Omega_{\bm r}),
\end{equation}
\begin{equation}
\int_0^{\infty} |R_{nl_i}|^2 r^2 dr=1.
\end{equation}
In our calculations we solve both the bound state and scattering
states numerically to obtain $R_{nl_i}$ and $R_{El}$. In the PWA,
the continuum electron is given by the plane waves
\begin{equation}
\Psi_{\bm k}({\bm r})=\frac{1}{(2\pi)^{3/2}}\exp(i{\bm k}\cdot{\bm
r}).
\end{equation}
where the interaction between the continuum electron and the target
ion in Eq.~(\ref{scat-Eq}) is completely neglected.

It is appropriate to make additional comments on the use of PWA for
describing the continuum electron since it is used to calculate the
dipole matrix elements in the SFA. Within this model, the target
structure enters only through the initial ground state. Thanks to
this approximation, the transition dipole moment is then given by
the Fourier transform of the ground state wavefunction weighted by
the dipole operator. Thus by performing inverse Fourier transform,
the ground state molecular wavefunction can be reconstructed from
the transition dipole moments. This forms the theoretical foundation
of the tomographic procedure used by Itatani {\it et al}
\cite{itatani-nature}. However, it is well-known that plane wave is
not accurate for describing continuum electrons at low energies
($\sim 20$ eV up to $\sim 0.5$ keV), that are typical for most HHG
experiments, and that all major features of molecular
photoionization in this energy range are attributable to the
property of continuum wavefunction instead of the ground state
wavefunction. In fact, the use of PWA in the SFA model is the major
deficiency of the SFA which has led to inaccuracies in the HHG
spectra, as shown earlier in \cite{atle08,H2+} and will also be
discussed in Sec.~III.

Note that so far we have  considered one-photon photoionization
process only. As mention earlier, the more relevant quantity to the
HHG process is its time-reversed one-photon photo-recombination
process.   The photo-recombination DCS can be written as
\begin{eqnarray}
\frac{d^2\sigma^R}{d\Omega_{\bm n}d\Omega_{\bm k}}=
\frac{4\pi^2\omega^3}{ck} \left|\bra\Psi_i|{\bm r}\cdot{\bm
n}|\Psi_{\bm k}^{+}\ket \right|^2. \label{CS-Rec}
\end{eqnarray}
In comparison with photoionization DCS in Eq.~(\ref{photo-DCS}),
apart from a different overall factor, the continuum state here is
taken as the outgoing scattering wave $\Psi_{\bm k}^+$ instead of an
incoming wave $\Psi_{\bm k}^-$. The partial wave expansion for
$\Psi_{\bm k}^+$ is
\begin{eqnarray}
\Psi_{\bm k}^{+}({\bm r})&=&\frac{1}{\sqrt{k}}
\sum_{l=0}^{\infty}\sum_{m=-l}^l i^l \exp[i(\sigma_l+\delta_l)]
R_{El}(r) Y_{lm}(\Omega_{\bm r}) Y_{lm}^*(\Omega_{\bm k}).
\end{eqnarray}
We note here the only difference is in the sign of the phase
$(\sigma_l+\delta_l)$ as compared to Eq.~(\ref{PW-expansion}). In
fact, the photoionization and photo-recombination DCS's are related
by
\begin{eqnarray}
\frac{d^2\sigma^R}{\omega^2d\Omega_{\bm n}d\Omega_{\bm k}}=
\frac{d^2\sigma^I}{k^2d\Omega_{\bm k}d\Omega_{\bm n}},
\end{eqnarray}
which  follows the principle of detailed balancing for the direct and
time-reversed processes \cite{landau}.

For clarity in the following we discuss the
photo-recombination in argon. Once the scattering wave is available,
the transition dipole can be calculated in the partial-wave
expansion as
\begin{eqnarray}
\bra\Psi_i|z|\Psi_{\bm k}^+\ket &=& \frac{1}{\sqrt{k}}\sum_{lm}
i^l e^{i(\sigma_l+\delta_l)}\bra R_{n l_i}|r|R_{E l}\ket \nonumber\\
& & \times\bra Y_{l_i m_i}|\cos\theta| Y_{l m}\ket
Y_{lm}^*(\Omega_{\bm k}).
\end{eqnarray}
Here the polarization direction ${\bm n}$ is assumed to be parallel
to $z$-axis. Using the relation
\begin{equation}
\cos\theta=\sqrt{\frac{4\pi}{3}}Y_{10}(\theta,\phi),
\end{equation}
the angular integration can be written as
\begin{equation}
\bra Y_{l_i m_i}|\cos\theta| Y_{l m}\ket=\sqrt{\frac{4\pi}{3}}\int
Y_{l_i m_i}^*(\theta,\phi)Y_{10}(\theta,\phi)Y_{l
m}(\theta,\phi)\sin\theta d\theta d\phi,
\end{equation}
which can also be expressed in terms of the Wigner 3j-symbol by
using
\begin{eqnarray}
\int Y_{l_1 m_1}(\theta,\phi)Y_{l_2m_2}(\theta,\phi)Y_{l_3
m_3}(\theta,\phi) \sin\theta d\theta
d\phi &=& \sqrt{\frac{(2l_1+1)(2l_2+1)(2l_3+1)}{4\pi}} \nonumber\\
 & & \times\left(
\begin{array}{ccc}
l_1 & l_2 & l_3\\
0 & 0 & 0
\end{array} \right)\left(
\begin{array}{ccc}
l_1 & l_2 & l_3\\
m_1 & m_2 & m_3
\end{array} \right).
\end{eqnarray}
From this equation, it is clear that only $m=m_i$ and $l=l_i-1$ and
$l=l_i+1$ contribute. This gives \cite{bethe}
\begin{eqnarray}
A_{l_im_i}^{lm}=\bra Y_{l_i m_i}|\cos\theta| Y_{l m_i}\ket=\left\{
\begin{array}{ll}
\sqrt{\frac{l_i^2-m_i^2}{(2l_i+1)(2l_i-1)}} & \quad \mbox{for
$l=l_i-1$} \\
\sqrt{\frac{(l_i+1)^2-m_i^2}{(2l_i+3)(2l_i+1)}} & \quad \mbox{for
$l=l_i+1$}
\end{array} \right.
\end{eqnarray}

As discussed in Sec.~II(A), we only need to consider the case of
recombination of electron back to Ar(3p$_0$), as electrons in $m=\pm
1$ states are not removed by tunnel ionization in the first step.
From the above equation we see that only $l=l_i-1=0$ (s-wave) and
$l=l_i+1=2$ (d-wave) contribute, with $A_{10}^{00}=1/\sqrt{3}$ and
$A_{10}^{20}=2/\sqrt{15}$, respectively. Furthermore, most
contribution to the HHG process comes from electrons moving along
laser's polarization direction. Since
\begin{eqnarray}
Y^*_{00}(0,0)=\sqrt{\frac{1}{4\pi}}, \nonumber \\
Y^*_{20}(0,0)=\sqrt{\frac{5}{4\pi}}, \nonumber
\end{eqnarray}
for ${\bm k}\parallel {\bm n}$ the transition dipole can be written
as
\begin{eqnarray}
\bra\Psi_i|z|\Psi_{\bm k}^+\ket &=& \frac{1}{\sqrt{3\pi k}}\left[
e^{i(\sigma_0+\delta_0)}\bra R_{3 1}|r|R_{E 0}\ket/2 \nonumber \right.\\
 & & \left. - e^{i(\sigma_2+\delta_2)}\bra R_{3 1}|r|R_{E 2}\ket
 \right]. \label{dipole-Ar}
\end{eqnarray}

Note that the transition dipole is intrinsically a complex number.
The dominant component is the d-wave. Thus when the real d-wave
radial dipole matrix element vanishes, the cross section will show a
minimum known as the Cooper minimum \cite{cooper}. Note that the
cross section does not go precisely to zero because of the
contribution from s-wave (the first term). We will analyze this
example in more detail in Sec.~III(B).

We comment that the calculation of transition dipole moment for
atomic targets presented above is based on the SAE. The validity of
such a model for describing photoionization has been studied in the
late 1960's, see the review by Fano and Cooper \cite{fano-rmp}. Such
a model gives an adequate description of the global energy
dependence of photoionization cross sections. To interpret precise
photoionization cross sections such as those carried out with
synchrotron radiation, advanced theoretical methods such as
many-body perturbation theory or R-matrix methods are needed. For
HHG, the returning electrons have a broad energy distribution as
opposed to the nearly monochromatic light from synchrotron radiation
light sources, thus the simple SAE can be used to study the global
HHG spectra.

\subsection{Calculation of transition dipoles for linear molecules}

Let us now consider photoionization of a linear molecule. For
molecules, the complication comes from the fact that the spherical
symmetry is lost and additional degrees of freedom are introduced.
The photoionization DCS in the body-fixed frame can be expressed in
the general form \cite{lucchese82}
\begin{equation}
\frac{d^2\sigma}{d\Omega_{\bm k}d\Omega_{\bm n}}= \frac{4\pi^2\omega
k}{c}\left|d_{{\bm k},{\bm n}}(\omega)\right|^2.
\end{equation}

For the emitted HHG component with polarization parallel to that of
the driving field, the only case to be considered is
 ${\bm k}\parallel{\bm n}$. To treat the dependence of the cross
section on the target alignment, it is convenient to expand the
transition dipole in terms of spherical harmonics
\begin{equation}
d_{{\bm k},{\bm n}}(\omega)=
\left(\frac{4\pi}{3}\right)^{1/2}\sum_{lm\mu}d_{lm\mu}(\omega)
Y^*_{lm}(\Omega_{\bm k})Y^*_{1\mu}(\Omega_{\bm n}).
\end{equation}
Here the partial-wave transition dipole is given by
\begin{equation}
d_{lm\mu}(\omega)=\langle\Psi_i|r_{\mu}|\Psi^-_{klm}\rangle,
\end{equation}
with $r_{\mu}=z$ for linear polarization.

In our calculations, we use an initial bound state obtained from the
MOLPRO code \cite{MOLPRO03} within the valence complete-active-space
self-consistent field (VCASSCF) method. The final state is then
described in a single-channel approximation where the target part of
the wavefunction is given by a valence complete active space
configuration interaction (VCASCI) wavefunction, obtained using the
same bound orbitals as are used in the wavefunction of the initial
state.
 The Schr\"odinger equation for the continuum
electron is
\begin{equation}
\left[-\frac{\nabla^2}{2}-\frac{1}{r}+\tilde{V}({\bm
r})-\frac{k^2}{2}\right]\phi_{{\bm k}}^-({\bm r})=0, \label{SECont}
\end{equation}
where $\tilde{V}({\bm r})$ is the short-range part of the electron
molecule interaction, which will be discussed below. Note that the
potential is not spherically symmetric for molecular systems. The
Schr\"odinger equation (\ref{SECont}) is then solved by using the
iterative Schwinger variational method \cite{lucchese82}. The
continuum wavefunction is expanded in terms of partial waves as
\begin{equation}
\phi_{{\bm k}}^-({\bm
r})=\left(\frac{2}{\pi}\right)^{1/2}\sum_{l=0}^{l_{p}}\sum_{m=-l}^l
\imath^l\phi^-_{klm}({\bm r})Y^*_{lm}(\Omega_{\bm k}),
\end{equation}
where an infinite sum over $l$ has been truncated at $l=l_{p}$. In
our calculations, we typically choose $l_{p}=11$. Note that our
continuum wavefunction is constructed to be orthogonal to the
strongly occupied orbitals. This avoids the spurious singularities
which can occur when scattering from correlated targets is
considered \cite{lucchese96}. We have used a single-center expansion
approach to evaluate all required matrix elements. That means that
all functions, including the scattering wavefunction, occupied
orbitals, and potential are expanded about a common origin, taken to
be the center of mass of the molecule, as a sum of spherical
harmonics times radial functions
\begin{equation}
F({\bm r})=\sum_{l=0}^{l_{max}}\sum_{m=-l}^l
f_{lm}(r)Y_{lm}(\theta,\phi).
\end{equation}
With this expansion, the angular integration can be done
analytically and all three-dimensional integrals reduce to a sum
of radial integrals, which are computed on a radial
grid. Typically, we use $l_{max}=60$ to 85.

Next we describe how the interaction potential, $\tilde{V}$ is constructed.
The electronic part of the Hamiltonian can be written as
\begin{equation}
H=\sum_{i=1}^N h(i)+\sum_{i<j}^N\frac{1}{r_{ij}},
\end{equation}
with
\begin{equation}
h(i)=-\frac{\nabla_i^2}{2}-\sum_{a}\frac{Z_a}{r_{ia}},
\end{equation}
where $Z_a$ are the nuclear charges and $N$ is the number of
electrons. In the single channel approximation used here, the
ionized state wavefunction $\Psi_{\bm k}$ is of the form
\begin{equation}
\Psi_{\bm k}=A\left( \Phi \phi_{\bm k} \right)
\label{SCCIwf}
\end{equation}
where $\Phi$ is the correlated $N-1$ electron ion core wavefunction,
$\phi_{\bm k}$ is the one-electron continuum wavefunction, and the
operator $A$ performs the appropriate antisymmetrization and spin
and spatial symmetry adaptation of the product of the ion and
continuum wavefunctions. The single-particle equation for the
continuum electron is obtained from
\begin{equation}
\langle\delta\Psi_{\bm k}|H-E|\Psi_{\bm k}\rangle=0,
\end{equation}
where $\delta\Psi_{\bm k}$ is written as in Eq.~(\ref{SCCIwf}), with
$\phi_{\bm k}$ replaced by $\delta\phi_{\bm k}$. By requiring this
equation to be satisfied for all possible $\delta\Psi_{\bm k}$ (or
$\delta\phi_{\bm k}$), one obtains a non-local optical potential
which can be written in the form of a Phillips-Kleinman
pseudopotential, $\tilde{V}$, \cite{lucchese88,lucchese93} as
written in Eq.~(\ref{SECont})

Note that ionization from a molecular orbital other than HOMO, say
HOMO-1, can be done in the same manner, except that the target state
$\Phi$ employed  in Eq.~(\ref{SCCIwf}) needs to be replaced by the
wavefunction for the corresponding excited ion state the corresponds
to ionization from the HOMO-1 orbital. Furthermore, the above
single-channel formalism can be extended to coupled-multichannel
calculations to account for additional electron correlation effects
\cite{lucchese95}. In this paper we limit ourselves to
single-channel calculations.

This single-center expansion approach has also been implemented for
non-linear targets in the frozen-core Hartree-Fock approximation
including the full non-local exchange potential \cite{natalense99}.
A somewhat simpler approach, the finite-element R-matrix (FERM3D) by
Tonzani \cite{tonzani} can also be employed to calculate transition
dipoles and photoionization cross sections. The FERM3D code is
especially well adapted for complex molecules. In FERM3D the
electrostatic potential is typically obtained from general {\it ab
initio} quantum chemistry software such as GAUSSIAN \cite{gaussian}
or GAMESS \cite{gamess} and the exchange potential is approximated
using a local density functional. A polarization potential is also
added to describe the long range attraction between the continuum
electron and the target ion. This code can also calculate ionization
from any occupied molecular orbitals.

\subsection{Description of strong-field ionization from the MO-ADK and the MO-SFA}

In the tunneling regime, the most successful general theories for
ionization from molecules are the MO-ADK \cite{moadk} and MO-SFA
\cite{faisal00,madsen05}. For our purpose of studying HHG process,
it is important to make clear distinction between the total
ionization (integrated over all emission directions) and the
(differential) ionization along the laser polarization direction.
The former is used in the description of the ground state depletion
[see Sec.~II(B)], while the latter is directly related to the
magnitude of the returning electron wave packet, which will be
described in Sec.~III(A). It is well-known that the SFA (or MO-SFA)
can give qualitatively good ATI spectra, but not the overall
magnitude \cite{chen06}, whereas the MO-ADK can give {\em total}
ionization rates. Strictly speaking, the MO-ADK only gives total
ionization, whereas the MO-SFA can also give differential rates.
Therefore in our calculations we use both theories. For N$_2$ and
O$_2$, the total ionization rates from the MO-ADK and the
(renormalized) MO-SFA agree very well \cite{hoang08}. However for
CO$_2$ the alignment dependent rates from the MO-ADK, the MO-SFA,
and the recent experiment \cite{NRC-exp} all disagree with each
other, with the MO-ADK predicting a peak near $30^{\circ}$, compared
to $40^{\circ}$ from the MO-SFA theory, and a very sharp peak near
$45^{\circ}$ from experiment \cite{NRC-exp}.

The MO-ADK theory is  described in details \cite{moadk}. Here we
will only mention briefly main equations in the MO-SFA theory. In
the MO-SFA model the ionization amplitude for a transition from a
bound state $\Phi_0({\bm r})$ to continuum is given by
\cite{lewenstein}
\begin{eqnarray}
\label{sfa}f({\bm p})=i\int_{-\infty}^{\infty} dt \langle {\bm
p}+{\bm A}(t)|{\bm r}\cdot{\bm E}(t)|\Psi_{0}\rangle \exp[-iS({\bm
p},t)],
\end{eqnarray}
where
\begin{eqnarray}
S({\bm p},t)=\int_{t} ^\infty dt'\left\{ \frac{[{\bm p}+{\bm
A}(t')]^{2}}{2}+I_{p}\right\},
\end{eqnarray}
with ${\bm p}$ the momentum of the emitted electron, $I_p$ the
binding energy of the initial state, and ${\bm A}(t)$ the vector
potential. In the SFA the effect of the core potential is totally
neglected in the continuum state, which is approximated by a Volkov
state
\begin{eqnarray}
\langle {\bm r}|{\bm p}+{\bm A}(t)\rangle=\frac{1}{(2
\pi)^{3/2}}\exp \left\{ i\left[ {\bm p}+{\bm A}(t) \right]\cdot{\bm
r} \right\}.
\end{eqnarray}
For the bound states we use the wavefunctions generated from the
{\it ab initio} quantum chemistry Gaussian \cite{gaussian} or Gamess
codes \cite{gamess}. In this regard we note that the use of these
wavefunctions in the MO-ADK might not be sufficiently accurate, as
the MO-ADK needs accurate  wavefunctions in the asymptotic region.

\subsection{Alignment distributions of molecules in laser fields}

When a molecule is placed in a short laser field (the pump), the
laser will excite a rotational wave packet (coherent superposition
of rotational states) in the molecule. By treating the linear
molecule as a rigid rotor \cite{seideman,ortigoso}, the rotational
motion of the molecule with initial state
$\Psi_{JM}(\theta,\phi,t=-\infty)=|JM\rangle$ evolves in the laser
field following the time-dependent Schr{\"o}dinger equation
\begin{equation}
i{\frac{\partial\Psi_{JM}(\theta,\phi,t)}{\partial t}}=[B{\bm
J}^2-{\frac{E(t)^2}{2}}(\alpha_{\parallel}\cos^2\theta
+\alpha_{\perp}\sin^2\theta)]\Psi_{JM}(\theta,\phi,t). \label{rotor}
\end{equation}
Here $E(t)$ is the laser electric field, $B$ is the rotational
constant, $\alpha_{\parallel}$ and $\alpha_{\perp}$ are the
anisotropic polarizabilities in parallel and perpendicular
directions with respect to the molecular axis, respectively. These
molecular properties for CO$_2$, O$_2$ and N$_2$ are given in
Table.~I. The above equation is then solved for each initial
rotational state $|JM\rangle$ using the split-operator method [see
Eq.~(\ref{split-operator})]. We assume the Boltzmann distribution of
the rotational levels at the initial time. With this assumption, the
time-dependent alignment distribution can be obtained as
\begin{equation}
\rho(\theta,t)=\sum_{JM}\omega_{JM}|\Psi_{JM}(\theta,\phi,t)|^2,
\end{equation}
where $\omega_{JM}$ is the weight according to the Boltzmann
distribution. Note that one needs to take proper account for the
nuclear statistics and symmetry of the total electronic
wavefunction. For example, in case of O$_2$ with total electron
wavefunction in $^3\Sigma_g^-$, only odd-$J$'s are allowed (see for
example \cite{herzberg}), whereas CO$_2$ with total electron
wavefunction in $^1\Sigma_g^+$ has only even-$J$'s. The angular
distribution or alignment does not depend on the azimuthal angle
$\phi$ in the frame attached to the pump laser field. The two
equations above allow the determination of the time-dependent
alignment distribution of the molecules in the laser field, as well
as the rotational revivals after the laser has been turned off. The
aligning laser is assumed to be weak enough so the molecules remain
in the ground state and no ionization occurs.

\begin{table}[ht]
\caption{Molecular properties for CO$_2$, O$_2$ and N$_2$. $B$ is
rotational constant, $\alpha_{\parallel}$ and $\alpha_{\perp}$ are
parallel and perpendicular polarizability, respectively. The data
are taken from \cite{hirschfelder,webbook}.} \centering
\begin{tabular}{c c c c}
\hline\hline Molecule & \quad $B$ (cm$^{-1}$) \quad & \quad
$\alpha_{\parallel}$
(\AA$^3$) \quad & \quad $\alpha_{\perp}$ (\AA$^3$) \quad \\[0.5ex]
\hline
CO$_2$ & 0.39 & 4.05 & 1.95 \\
O$_2$  & 1.4377 & 2.35 & 1.21 \\
N$_2$  & 1.989  & 2.38 & 1.45 \\[1ex]
\hline
\end{tabular}
\end{table}

Once the angular distribution is obtained, the (complex) induced
dipole for emission of photon energy of $\omega$ can be calculated
by adding coherently the weighted contribution from different
alignments by

\begin{equation}
\overline{D}(\omega,t)=2\pi\int_0^{\pi} D(\omega,\theta)
\rho(\theta,t)\sin\theta d\theta,
\end{equation}
if the pump and probe laser polarizations are parallel. Here we
assume that rotational motion during the femtosecond probe pulse is
negligible, which should be valid for the molecules under
consideration with typical rotational period of few picoseconds.


If the polarizations of the pump and probe lasers are not the same,
the theoretical treatment is rather cumbersome as the cylindrical
symmetry is lost. This case has been discussed by Lein {\it et al}
\cite{lein-jmo05}. Here we briefly describe the main results. Assume
that the pump and probe laser pulses propagate collinearly and
$\alpha$ is the angle between the two polarization directions. Let
$\theta$ ($\theta'$) and $\phi$ ($\phi'$) be the polar and azimuthal
angles of the molecular axis in the frame attached to the pump
(probe) field. These angles are related by

\begin{equation}
\cos\theta=\cos\theta'\cos\alpha+\sin\theta'\sin\alpha\cos\phi'.
\end{equation}
The alignment distribution in the ``probe'' frame is

\begin{equation}
\rho(\alpha,\theta',\phi',t)=\rho(\theta(\alpha,\theta',\phi'),t).
\end{equation}
For the emitted HHG with polarization parallel to that of the probe
laser, the induced dipole can then be obtained from
\begin{equation}
\overline{D}_{\parallel}(\omega,\alpha,t)=\int_0^{\pi}\int_0^{2\pi}
D_{\parallel}(\omega,\theta')\rho(\alpha,\theta',\phi',t)\sin\theta'd\theta'd\phi'.
\end{equation}
And for the perpendicular component
\begin{eqnarray}
\overline{D}_{\perp}(\omega,\alpha,t)&=&\int_0^{\pi}\int_0^{2\pi}
D_{\perp}(\omega,\theta',\phi')\rho(\alpha,\theta',\phi',t)\sin\theta'd\theta'd\phi'
\nonumber\\
&=& \int_0^{\pi}\int_0^{2\pi}
D_{\perp}(\omega,\theta',\phi'=0)\rho(\alpha,\theta',\phi',t)
\sin\theta'\cos\phi'd\theta'd\phi'.
\end{eqnarray}

\section{Quantitative Rescatering theory}

In this section we provide formulation of the quantitative
rescatering theory (QRS) and theoretical evidence in supporting its
validity and improvements over the very popular SFA (or the
Lewenstein model). Particular attention is given to the
photo-recombination differential cross sections (DCS) and phases,
which are used to illustrate the nature of the improvements and will
also be used in the next section to simulate data for comparison
with experiments.

\subsection{Description of the QRS}
Within the QRS as applied to HHG process, induced dipole
$D(\omega,\theta)$ and its phase $\varphi(\omega,\theta)$ for a
molecule aligned with an angle $\theta$ with respect to the laser
polarization can be written as
\begin{equation}
D(\omega,\theta)=W(E,\theta)d(\omega,\theta), \label{QRS-model-1}
\end{equation}
or more explicitly
\begin{equation}
|D(\omega,\theta)|e^{i\varphi(\omega,\theta)}=|W(E,\theta)|e^{i\eta}
|d(\omega,\theta)|e^{i\delta(\omega,\theta)}, \label{QRS-model-2}
\end{equation}
where $d(\omega,\theta)$ and $\delta(\omega,\theta)$ are the
``exact'' transition dipole and its phase defined in sections II(C)
and (D). The quantity $|W(E,\theta)|^2$ describes the flux of the
returning electrons, which we will call a ``wave packet", with
$\eta(E,\theta)$ being its phase. Electron energy $E$ is related to
the emitted photon energy $\omega$ by $E=\omega-I_p$, with $I_p$
being the ionization potential of the target. Clearly the HHG signal
$S(\omega,\theta)\sim |D(\omega,\theta)|^2$ and $W(E,\theta)$ depend
on the laser properties. On the other hand, $d(\omega,\theta)$ is
the property of the target only. Equation (\ref{QRS-model-1}) can be
also seen as the definition of the wave packet, assuming the induced
dipole and transition dipole are known. Since the returning wave
packet is an important concept in the QRS theory, let us write it
down explicitly
\begin{equation}
W(E,\theta) = \frac{D(\omega,\theta)}{d(\omega,\theta)}.
\label{wave-packet}
\end{equation}

The validity of Eq.~(\ref{QRS-model-1}) on the level of {\em
amplitudes} has been shown in Morishita {\it et al.} \cite{toru}
using HHG spectra calculated by solving the TDSE for atoms.
Indications for the validity of this factorization have also been
shown for rare gas atoms by Levesque {\it et al.} \cite{levesque07}
and for N$_2$ and O$_2$ molecules \cite{hoang07}, where the HHG
spectra were calculated using the SFA model with the continuum
electron being treated in the plane wave approximation.

In the tunneling regime, most electrons will be driven along the
laser polarization direction. In fact, semi-classical treatment
shows that most contribution to the HHG process is coming from
electron released to the continuum and returning to the target ion
along the laser polarization direction \cite{lewenstein}. In the
notations adopted in Sec.~II(C) and (D), with ${\bm n}$ and ${\bm
n'}$ being the directions of the driving laser and HHG
polarizations, respectively, and ${\bm k}$ the electron momentum, we
have
\begin{eqnarray}
d(\omega,\theta)\equiv d_{{\bm k},{\bm n'}}(\omega,\theta) \mbox{
 with } \left\{
\begin{array}{ll}
{\bm k}\parallel{\bm n}\parallel{\bm n'} & \quad
\mbox{for $D_{\parallel}(\omega,\theta)$} \\
{\bm k}\parallel{\bm n}\perp{\bm n'} & \quad \mbox{for
$D_{\perp}(\omega,\theta)$}.
\end{array} \right. \label{polar}
\end{eqnarray}
In this paper, we limit ourselves to the parallel component of HHG,
and therefore the subscripts are omitted in the notations.

The usefulness of   Eq.~(\ref{QRS-model-1}) is two-fold. First, the
factorization allows us to separate in HHG process the effect of the
laser field (on the returning electron wave packets) and the
influence of target structure (transition dipole and its phase).
From practical point of view, this implies that one can carry out
calculations for each factor separately. It is particularly
important to be able to use the vast knowledge of the molecular
photoionization/photo-recombination processes, which have been
accumulated over the last few decades. As for the wave packets, it
will be shown that although the HHG spectra from the SFA are not
quite accurate,  the extracted wave packets are reasonably good.
This offers a simple and efficient way to calculate the wave
packets. Second, the wave packets can also be shown to be largely
independent of the targets. By that we mean that the shape of the
wave packet as a function of electron energy depends only on laser
parameters, for targets with similar ionization potentials. In other
words, the wave packets can be written as
\begin{equation}
|W(E,\theta)|^2=N(\theta, ...)\times|\tilde{W}(E)|^2, \label{normal}
\end{equation}
where $N(\theta, ...)$ is the ionization probability for the
emission along the laser polarization direction, which depends on
the alignment angle, symmetry of the HOMO, and other parameters. The
overall factor $N(\theta, ...)$ does not change the shape of
$\tilde{W}(E)$. This offers another way to obtain the wave packets
by using a reference atom, for which numerical solution of the TDSE
is relatively simpler than that of a molecule under consideration.
Regarding the factor $N(\theta, ...)$, it is well understood for
many molecular systems based on the MO-ADK theory \cite{moadk} and
MO-SFA [see Sec.~II(E)], and the numerical solution of the TDSE for
simple systems \cite{kamta05,madsen07,telnov07,jiang08}. Note that,
the numerical solution of the TDSE for ionization rate is much less
computational demanding than for calculating HHG spectra.

From the above general discussion, let us now be more specific about
the two ways of obtaining the wave packets and HHG spectra within
the QRS model. For a given target, we can obtain the wave packets
from the SFA. If the laser pulse is given, one first uses the SFA to
calculate $D^{SFA}(t,\theta)$ and its Fourier transform
$D^{SFA}(\omega,\theta)$ by using Eq.~(11). Taking into account that
the PWA is used in the SFA, we have
\begin{equation}
W^{SFA}(E,\theta) =
\frac{D^{SFA}(\omega,\theta)}{d^{PWA}(\omega,\theta)}.
\label{wp-SFA}
\end{equation}
It will be shown that this wave packet agrees reasonably well with
the ``exact'' wave packet $W(E,\theta)$ obtained from solving the
TDSE. Once the wave packet is obtained, the induced dipole can be
calculated by
\begin{eqnarray}
D^{QRS1}(\omega,\theta)&=&W^{SFA}(E,\theta)d(\omega,\theta) \nonumber \\
&=&
\frac{d(\omega,\theta)}{d^{PWA}(\omega,\theta)}D^{SFA}(\omega,\theta).
\label{QRS1}
\end{eqnarray}
Since the  photo-recombination  cross section is proportional to
modulus square of the transition dipole [see Eqs.~(14), (15), (25),
and (33)], the HHG yield can be written as
\begin{eqnarray}
S^{QRS1}(\omega,\theta) &=& \left|\frac{d(\omega,\theta)}
{d^{PWA}(\omega,\theta)}\right|^2S^{SFA}(\omega,\theta) \nonumber \\
&=& \frac{\sigma(\omega,\theta)}{\sigma^{PWA}(\omega,\theta)}
S^{SFA}(\omega,\theta),
\end{eqnarray}
where $\sigma$ and $\sigma^{PWA}$ are the short-hand notations for
the ``exact'' and PWA differential   photo-recombination cross
sections, respectively. For simplicity, we have used QRS1 to denote
this version of the QRS model. Since the wave packets are obtained
from the SFA, this method was called the scattering-wave based
strong-field approximation (SW-SFA) in our previous papers
\cite{atle08,H2+}. This method has a very simple interpretation: the
QRS corrects the inaccuracies in the HHG yield from SFA by a simple
scaling factor, equal to the ratio of the exact and approximate
photo-recombination cross sections. It also adds the exact
transition dipole phase to the harmonic phase. The success of this
version of the QRS lies in the fact that the SFA describes the
electron wave packet quite accurately. Recall in the three-step
model, in the second step the electron ``roams'' well outside the
target ion most of the time before being driven back to recollide
with the ion. During this excursion, the electron motion is governed
mostly by the laser field, which is well described by the SFA. Thus,
in this version only the transition dipole moment is corrected.

The second method of obtaining the wave packet for the QRS is to use
a reference atom with a similar ionization potential. For the
reference atom, we can perform the TDSE calculation. Using the idea
of the QRS, one can obtain the wave packet from
\begin{equation}
W^{ref}(E) = \frac{D^{ref}(\omega)}{d^{ref}(\omega)}. \label{wp-ref}
\end{equation}
The power of this method stems from the fact that effect of the
target potential on the wave packet is included to some extent in
the second step of the three-step model (see the paragraph above),
when electron is quite far from the ion core and sees mostly the
long-range Coulomb tail. This method has the advantage in improving
the accuracy of the phase of the HHG induced dipole, but it is much
more time consuming. Combining with Eq.~(67), the wave packet for
the molecular system of interest can then be written as
\begin{eqnarray}
W^{QRS2}(E,\theta)&=&\left(\frac{N(\theta)}{N^{ref}}\right)^{1/2} W^{ref}(E)e^{i\Delta\eta} \nonumber
\\
&=& \left(\frac{N(\theta)}{N^{ref}}\right)^{1/2}
\frac{D^{ref}(\omega)}{d^{ref}(\omega)}e^{i\Delta\eta},
\end{eqnarray}
where $N(\theta)$ and $N^{ref}$ are the ionization probability for
electron emission along the laser polarization direction from the
molecule and reference atom, respectively. $\Delta\eta$ is
introduced to account for the phase difference between the two wave
packets. This phase difference will be shown to be nearly
independent of energy. The induced dipole and HHG spectra can then
be written as
\begin{eqnarray}
D^{QRS2}(\omega,\theta)&=&W^{QRS2}(E,\theta)d(\omega,\theta) \nonumber \\
&=& \left(\frac{N(\theta)}{N^{ref}}\right)^{1/2}
\frac{d(\omega,\theta)}{d^{ref}(\omega)}e^{i\Delta\eta}D^{ref}(\omega),
\label{QRS2}
\end{eqnarray}
and
\begin{eqnarray}
S^{QRS2}(\omega,\theta) = \frac{N(\theta)}{N^{ref}}
\frac{\sigma(\omega,\theta)}{\sigma^{ref}(\omega)} S^{ref}(\omega).
\end{eqnarray}

This version of the QRS, called QRS2 above, also has important
implications. It reflects the fact that in the tunneling regime the
returning electron wave packet has nearly identical shape (momentum
distribution) for all targets, except for an overall factor
accounting for the differences in ionization rates. In practice for
the reference atom we use a scaled hydrogen with the effective
nuclear charge chosen such that it has the same $1s$ binding energy
as the molecule under consideration. Experimentally, one can replace
the scaled atomic hydrogen with an atomic target of comparable
ionization potential. The wave packet obtained from the reference
atom also has the advantage that it avoids the spurious singularity
often seen in the wave packet obtained from the SFA. Such
singularity occurs since the transition dipole calculated from PWA
usually goes to zero at some photon energy, see Eq.~(\ref{QRS1}).
Both versions of the QRS have been used. In general, the one using
the reference atom is more accurate but takes much longer time since
the wave packets are obtained from the TDSE for atoms. Interested
reader is also referred to our recent papers
\cite{toru,atle08,H2+,atle09} for other evidences and applications
of the QRS to HHG processes. We note that the factorization can be
approximately derived analytically based on the SFA
\cite{atle-unpub} and within the zero-range potential approach
\cite{frolov09}.

\subsection{Atomic photo-recombination cross sections and phases}

In this subsection we will consider an example of
photo-recombination of Ar(3p$_0$), which will be used in the QRS
calculation in the next subsection. We will use a single-active
electron model with a model potential, as detailed in Sec.~II(A) and
(C). We have found that the position of the Cooper minimum is quite
sensitive to the form of the model potential. In our earlier paper
\cite{toru,atle08} we used the model potential of Tong and Lin
\cite{tong-jpb05}, which shows a Cooper minimum near 42 eV. To have
more realistic simulations for HHG spectra in this paper we use a
model potential, suggested by Muller \cite{muller}, which has been
shown to be able to reproduce the ATI spectra comparable with
experiments \cite{muller2}. This model potential has also been used
quite recently to simulate HHG experiments by Minemoto {\it et al}
\cite{minemoto08} and W\"orner {\it et al} \cite{woerner09}. In
Fig.~1(a) and 1(b) we compare the differential photo-recombination
cross sections and dipole phases obtained from Eqs.~(\ref{CS-Rec})
and (\ref{dipole-Ar}) with these two model potentials. We also plot
here the results from the PWA, with the ground state wavefunction
from Muller potential, which are almost identical with the PWA
results from Tong and Lin potential (not shown). Clearly, in the
energy range shown in the figure the PWA result deviates
significantly from the two more accurate results. Furthermore, the
Cooper minimum obtained from the potential by Muller occurs near 50
eV, in a better agreement with experiments, although the minimum is
somewhat more shallow than that from Tong and Lin model potential.
This is not surprising since the scattering waves are known to be
more sensitive to the details of the potential than the bound
states, which are localized near the target core. The dipole phases
obtained from both potentials show dramatic jumps of about 2 radians
near the Cooper minima. Note that the dipole phase from the PWA
shows a phase jump by $\pi$ at the ``Cooper minimum'' near 21 eV.
The failure of PWA for describing photo-recombination cross section
and phase at low energies is well-known. Other examples can be found
in Ref.~\cite{atle08} for rare gas atoms and in Ref.~\cite{H2+} for
H$_2^+$.

\begin{figure}
\mbox{\rotatebox{0}{\myscaleboxd{
\includegraphics{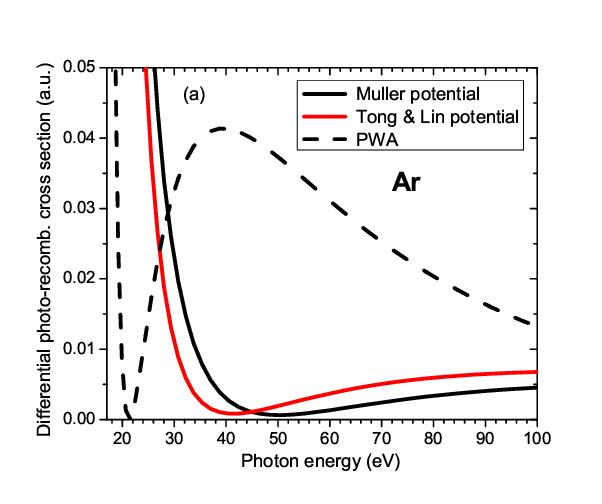}}}}
\mbox{\rotatebox{0}{\myscaleboxd{
\includegraphics{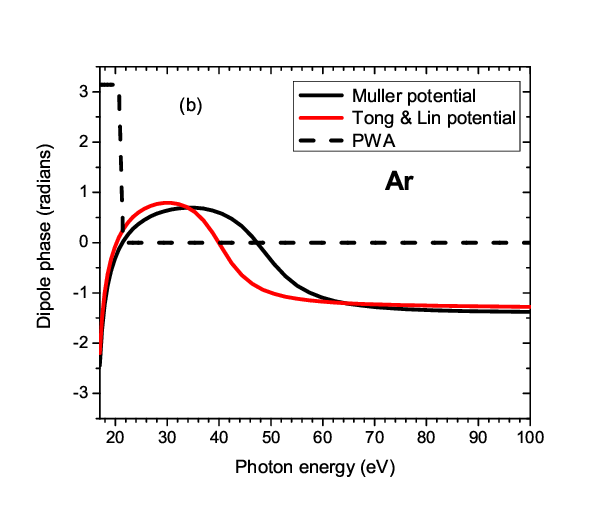}}}}
\caption{(Color online) Comparison of the differential
photo-recombination cross sections (a) and dipole phases (b) from
different models for Ar.}
\end{figure}

The simplicity of the model potential approach allows us to
establish the validity of the QRS for atoms, by using the same
potential in the TDSE and in the calculation of the
photo-recombination  transition dipole (see next subsection). In
order to compare with experiments, one can go beyond the model
potential approach by using different schemes to account for
many-electron effect, i.e., final state and initial state
correlation in photoionization (see, for example,
\cite{starace,lin74,amusia71}). It is much harder to do so within
the TDSE approach.

\subsection{QRS for atomic targets: example of Argon}

By using the photo-recombination cross section and dipole phase
shown in the previous subsection, we are now ready to discuss the
results from the QRS and assess its validity on an example of HHG
from Ar(3p$_0$). We emphasize that in this paper we use a model
potential proposed by Muller \cite{muller} in order to get the
Cooper minimum position comparable with experiments.

In Fig.~2(a) we show the comparison of the returning electron wave
packets from both versions of the QRS with the ``exact'' wave
packet, extracted from the solution of the TDSE for Ar by using
Eq.~(65). In the QRS2 the effective charge $Z_{eff}=1.0763$ is
chosen such that the ionization potential from $1s$ state is 15.76
eV, the same as for Ar(3p$_0$). We use a 800-nm wavelength laser
pulse with 8-cycle duration (8 fs FWHM) and peak intensity of
$2.5\times 10^{14}$ W/cm$^2$. One can see very good agreement
between QRS2 and the exact result over a very broad range of energy
when the wave packet is extracted from the scaled atomic hydrogen.
The result from the  QRS1 (shifted vertically for clarity) with the
wave packet from the SFA is also in reasonable good agreement with
the exact one.

\begin{figure}
\mbox{\rotatebox{0}{\myscaleboxd{
\includegraphics{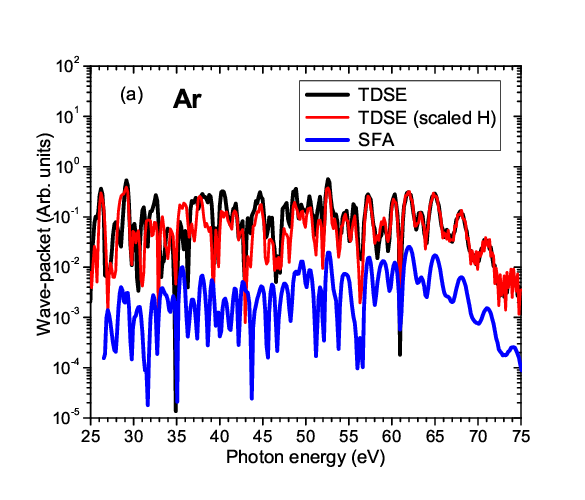}}}}
\mbox{\rotatebox{0}{\myscaleboxd{
\includegraphics{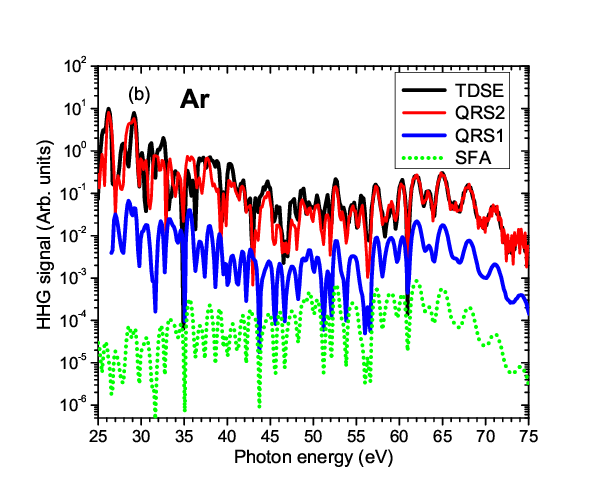}}}}
\caption{(Color online) (a) Comparison of the returning electron
``wave packets'' extracted from numerical solutions of the TDSE for
Ar and scaled hydrogen and from the SFA for Ar. (b) Comparison of
the HHG yields obtained from numerical solutions of the TDSE, QRS
and SFA for Ar. Data have been shifted vertically for clarity.
8-cycle laser pulse with peak intensity of $2.5\times 10^{14}$
W/cm$^2$, 800-nm wavelength is used.}
\end{figure}

Fig.~2(b) shows comparison of the HHG yields from the TDSE, QRS1,
QRS2 and the SFA. The data from QRS1 and SFA have also been shifted
vertically for clarity. Clearly, the QRS results agree quite well
with the TDSE, whereas the SFA result deviates strongly in the lower
plateau. The signature of the Cooper minimum near 50 eV in the HHG
spectra is quite visible. Note that the minimum has shifted as
compared to the results reported in Ref.~\cite{atle08}, which used a
model potential suggested by Tong and Lin \cite{tong-jpb05}. The
results in Fig.~2(b) clearly demonstrate the good improvement of the
QRS over the SFA in achieving better agreement with the TDSE
results.

\begin{figure}
\mbox{\rotatebox{0}{\myscaleboxd{
\includegraphics{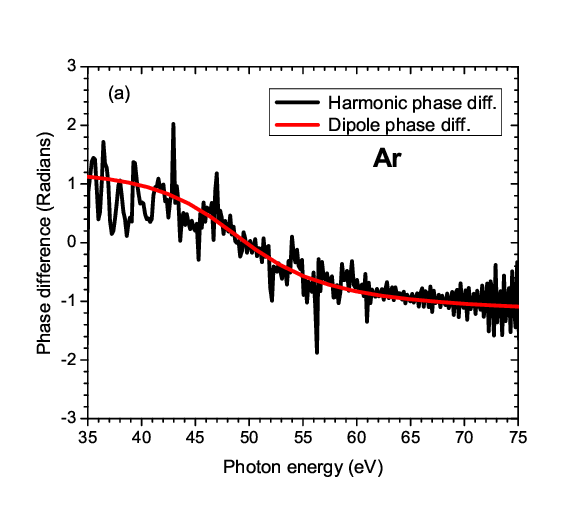}}}}
\mbox{\rotatebox{0}{\myscaleboxd{
\includegraphics{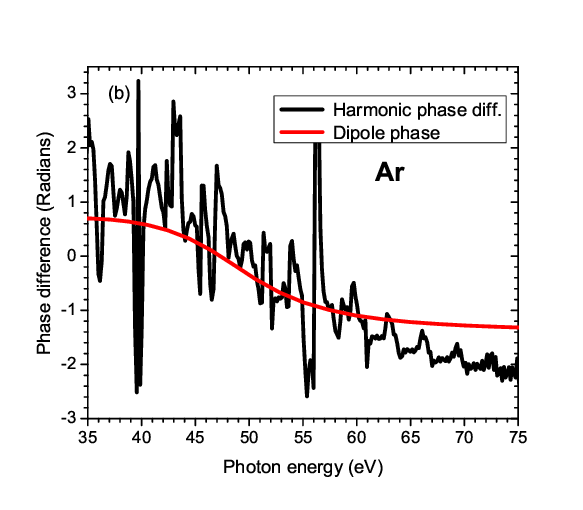}}}}
\caption{(Color online) (a) Comparison of extracted harmonic phase
difference $\Delta\varphi$ between Ar and scaled hydrogen as a
function of emitted photon energy with the photo-recombination
transition dipole phase difference $\Delta\delta$. The results are
obtained from numerical solutions of the TDSE. (b) Same as for (a),
but SFA for Ar is used instead of a scaled hydrogen. Laser
parameters are the same as for Fig.~2.}
\end{figure}

Next we examine harmonic phase, or induced dipole phase $\varphi$
[see Eq.~(\ref{QRS-model-2})]. Since harmonic phase changes quickly
from one order to the next (except in the harmonic cut-off), it is
instructive to compare phases from different targets. For that
purpose we calculate the harmonic phase difference $\Delta
\varphi=\varphi-\varphi^{ref}$ of Ar from its reference scaled
hydrogen. The result, shifted by 1.9 radians, is shown as the solid
black curve in Fig.~3(a), at an energy grid with step-size of
$0.1\times\omega_0$, where $\omega_0$ is the photon energy of the
driving laser ($1.55$ eV). A striking feature is that this curve
agrees very well with the transition dipole phase difference
$\Delta\delta=\delta-\delta^{ref}$, shown as the red line. Since the
transition dipole phase $\delta^{ref}$ from the reference scaled
hydrogen is very small (about $0.2$ radians) in this range of
energy, $\Delta\delta$ looks quite close to $\delta$ as well [see
Fig.~1(b)]. Note the phase jump near the Cooper minimum at 50 eV is
also well reproduced. This clearly demonstrates the validity of the
QRS model in Eqs.~(\ref{QRS-model-2}) and (\ref{QRS2}) and with
respect to the phase. The shift by 1.9 radians can be attributed to
the phase difference of the two wave packets $\Delta\eta$, which is
nearly energy independent.

Above result is based on the numerical solution of the TDSE for Ar
and scaled hydrogen. Similarly one can compare the harmonic phase
difference between the TDSE and SFA results $\Delta
\tilde{\varphi}=\varphi-\varphi^{SFA}$ with that of the transition
dipole phase $\Delta \tilde{\delta}=\delta-\delta^{PWA}$. Note that
$\delta^{PWA}=0$ [see Fig.~2(b)]. Clearly, if the SFA and the PWA
were exact, one would have $\Delta \tilde{\varphi}=\Delta
\tilde{\delta}=0$. The results, presented in Fig.~3(b), show that
the phase missed in the SFA is quite close to the transition dipole
phase. The QRS1 therefore corrects the inaccuracy in the SFA phase
by adding the phase from the transition dipole [see
Eq.~(\ref{QRS1})]. By comparing Figs.~3(a) with 3(b), we conclude
that the phase from QRS2 is more accurate than QRS1. This is
expected since the QRS2 includes partially the effect of
electron-core interaction during the propagation in the continuum,
as explained in III(A).

\begin{figure}
\mbox{\rotatebox{0}{\myscaleboxc{
\includegraphics{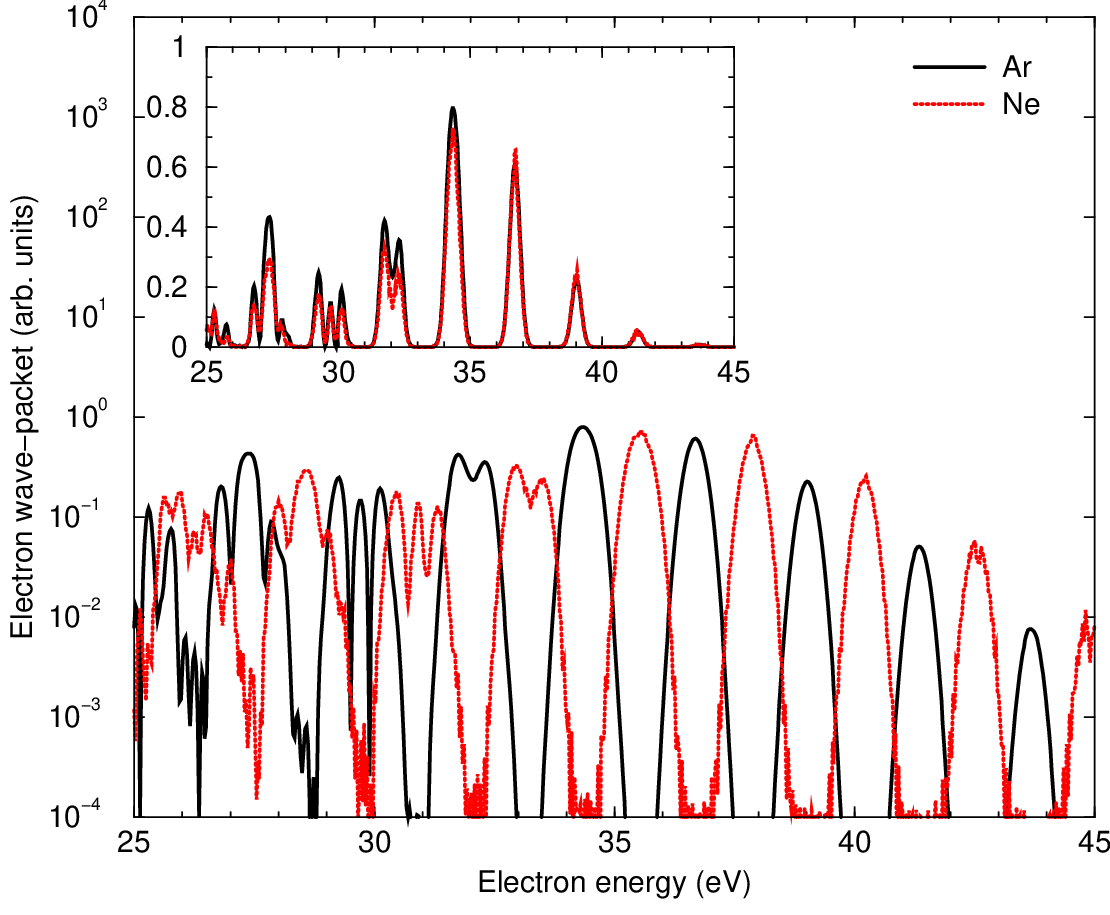}}}}
\caption{(Color online) The electron wave packets, extracted from
the HHG spectra for Ne and Ar under the same laser field. The inset
shows the wave packets in the linear scale with the Ne data shifted
horizontally by -1.2 eV. The HHG are obtained within the Lewenstein
model with 1064 nm laser, peak intensity of $2\times 10^{14}$
W/cm$^2$, duration of 50 fs.} \label{Ar_Ne}
\end{figure}

Finally,  we remark that the use of a reference atom with nearly
identical ionization potential is not necessary.  We will show here
this requirement can be relaxed. As the wave packets obtained within
the SFA agree reasonably well with the TDSE results, for our purpose
we only use the SFA in the subsequent analysis. In Fig.~4, we show
the comparison of the wave packet from Ne ($I_p=21.56$ eV) and Ar
($I_p=15.76$ eV) in the same 1064 nm laser pulse ($\omega=1.166$ eV)
with peak intensity of $2\times 10^{14}$ W/cm$^2$, duration (FWHM)
of 50 fs. Note that the wave packets are now plotted as functions of
{\em electron} energy, instead of the photon energy as in Fig.~2(b).
Clearly, the two wave packets lie nicely within a common envelope.
This can be seen even more clearly in the inset, where the data are
plotted in linear scale with the Ne data shifted horizontally by
$-1.2$ eV. We note that the small details below 33 eV also agree
well. This conclusion is also confirmed by our calculations with
different laser parameters and with other atoms. This supports that
the independence of the wave packet on the target structure can also
be extended to systems with different ionization potentials. This
fact can be useful in comparing experimental data for different
targets. Note that the small shift of -1.2 eV is caused by the fact
that the difference in the ionization potentials (5.8 eV) is
incommensurate with the energy of two fundamental photons (2.3 eV).

\subsection{Molecular photoionization cross sections and phases}

Photoionization cross section and transition dipole phase from a
linear molecule depend on both photon energy and angle between
molecular axis and laser polarization direction. Since most
interesting features of HHG from molecules are from molecules that
have been pre-aligned by a pump laser pulse, it is more appropriate
to present them as functions of fixed alignment angle for fixed
energies. As noted before, one can use either photoionization or
photo-recombination cross sections and phases in the QRS. Here we
choose to use photoionization as it is in general more widely
available theoretically and experimentally. In this paper we are
only interested in the HHG with polarization parallel to that of the
driving laser. That limits the differential cross sections to that
case of ${\bm k}\parallel{\bm n}\parallel{\bm n}'$ [see
Eq.~(\ref{polar})].

\begin{figure}
\mbox{\rotatebox{0}{\myscaleboxd{
\includegraphics{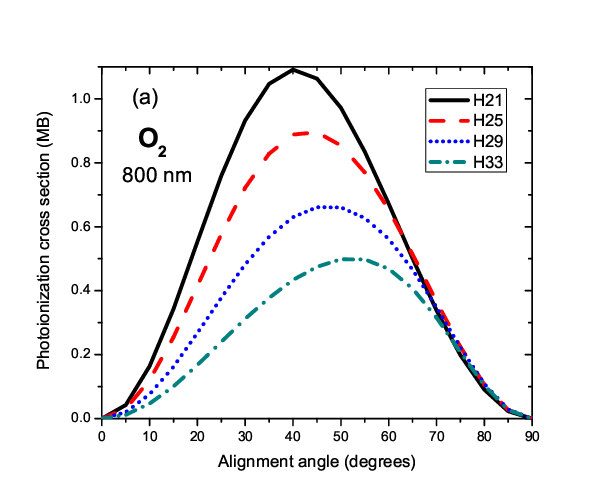}}}}
\mbox{\rotatebox{0}{\myscaleboxd{
\includegraphics{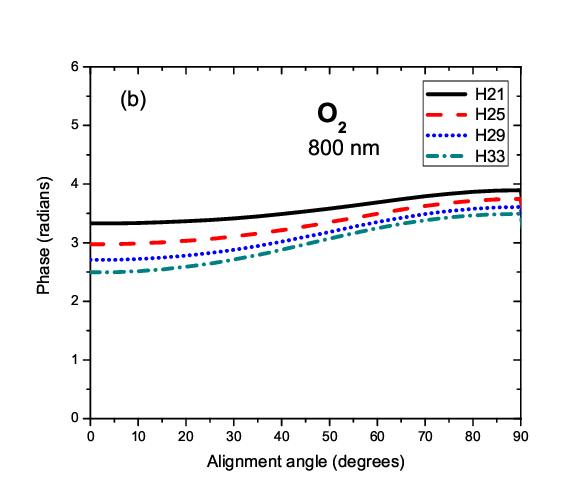}}}}

\mbox{\rotatebox{0}{\myscaleboxc{
\includegraphics{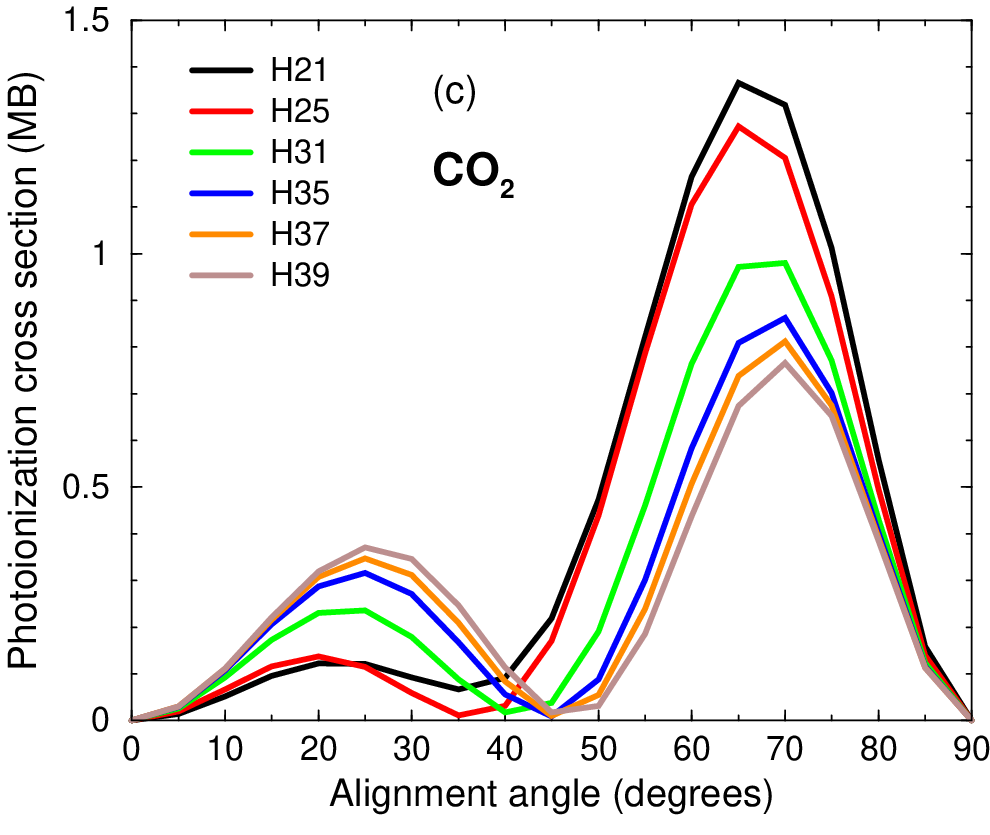}}}}
\mbox{\rotatebox{0}{\myscaleboxc{
\includegraphics{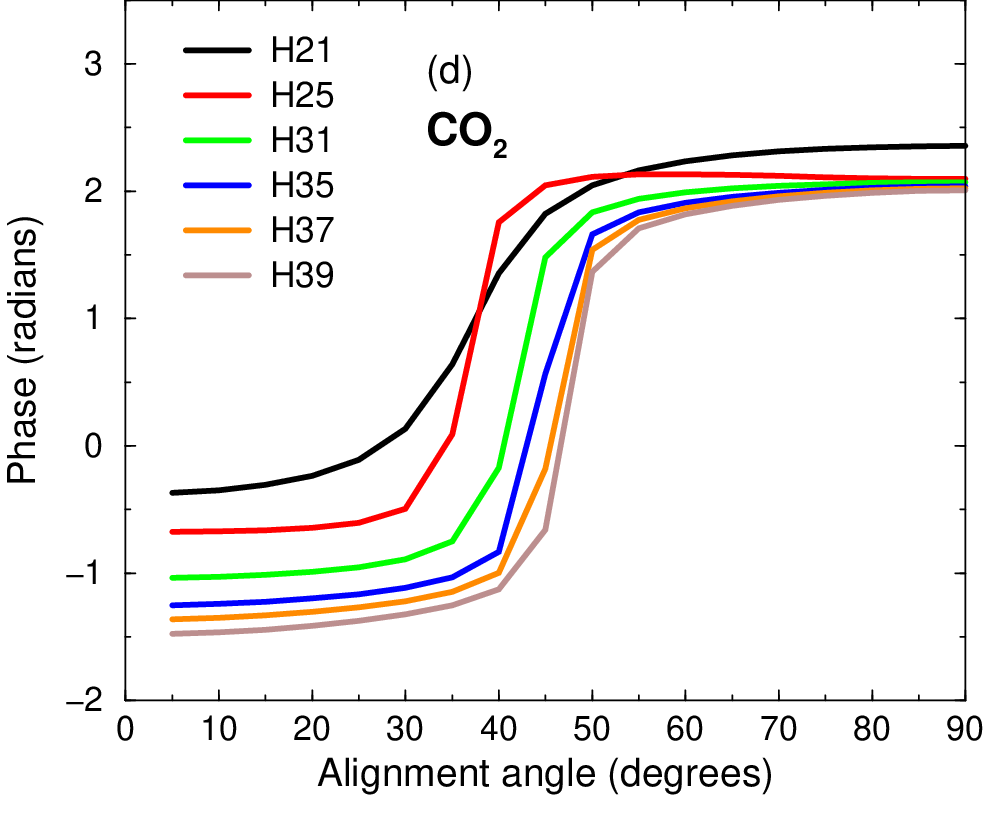}}}}

\mbox{\rotatebox{0}{\myscaleboxc{
\includegraphics{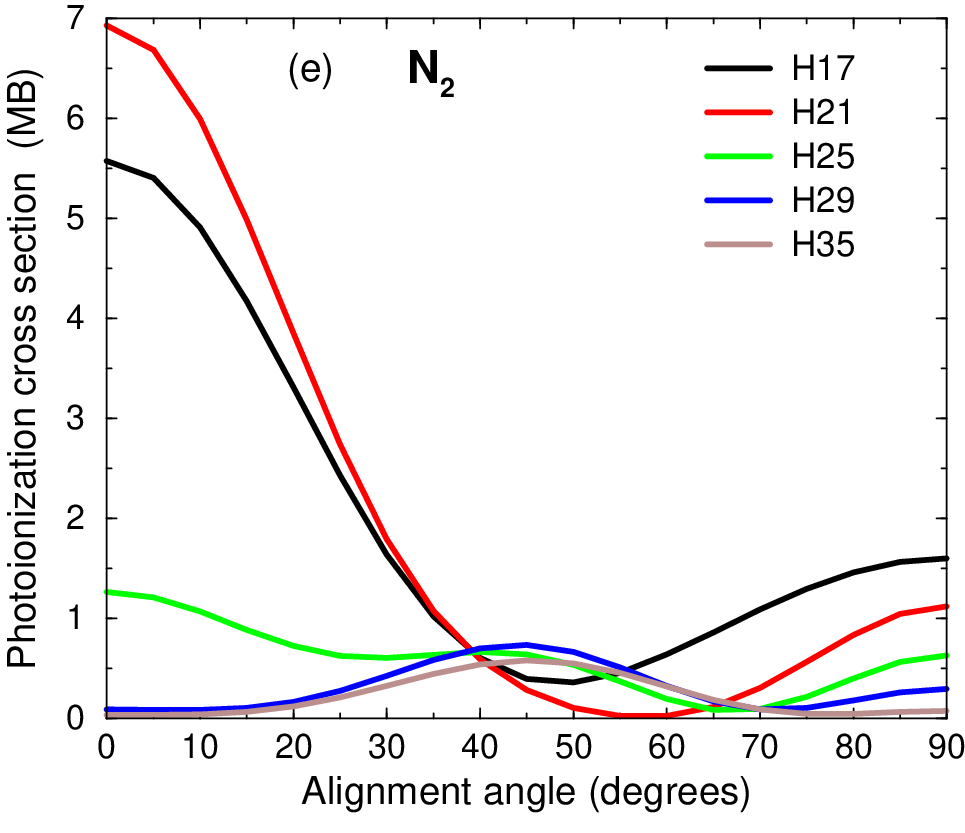}}}}
\mbox{\rotatebox{0}{\myscaleboxc{
\includegraphics{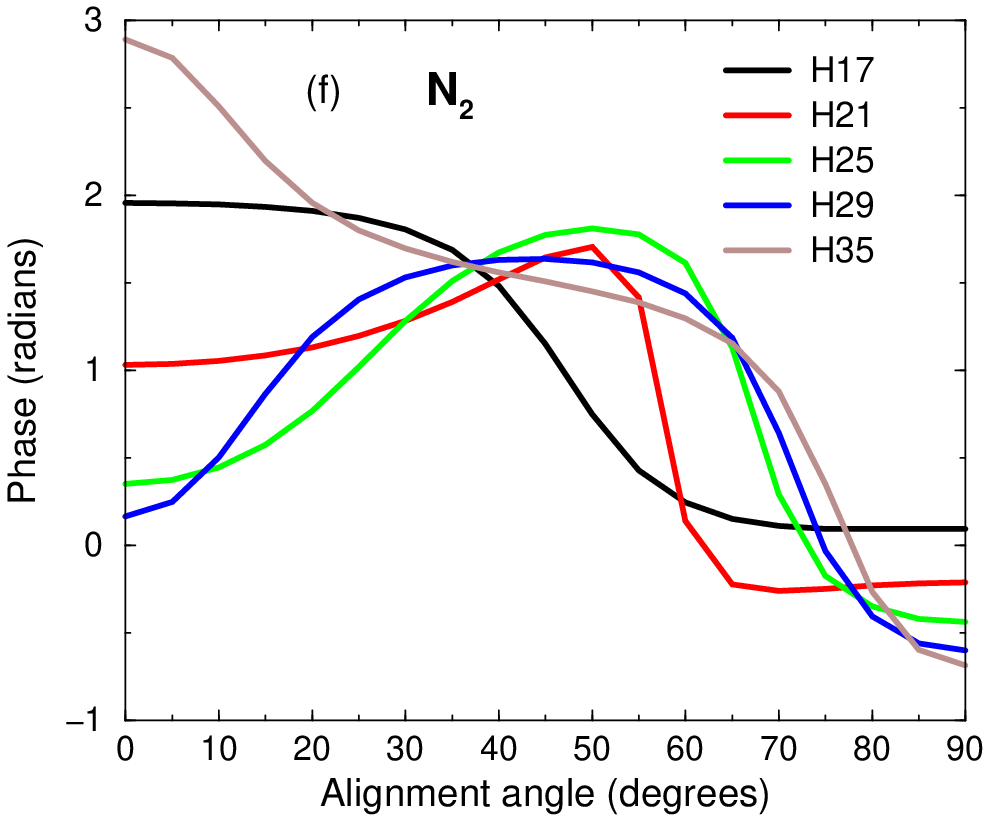}}}}

\caption{(Color online) Diffrential photoionization cross sections
and transition phases for O$_2$, CO$_2$, and N$_2$, as functions of
the alignment angle. The photon energies are expressed in units of
harmonic orders for 800-nm laser.}
\end{figure}

The photoionization DCS and phase from O$_2$, CO$_2$, and N$_2$ are
presented in Fig.~5(a-f) for some fixed energies as functions of
angle between molecular axis and laser polarization direction. For
convenience we express energy in units of photon energy of 800-nm
laser. Let us first examine O$_2$, shown in Fig.~5(a) and (b). For a
fixed energy the cross section vanishes at $\theta=0$ and $\pi/2$
due to the $\pi_g$ symmetry of the HOMO and the dipole selection
rule for the final state and has a peak near $45^{\circ}$.  As
energy increases the peak slightly shifts to a larger angle and the
cross section monotonically decreases. The phases behave quite
smoothly and change only within about 1 radian for all energies
considered. For CO$_2$, the cross section also vanishes at
$\theta=0$ and $\pi/2$ due to the $\pi_g$ symmetry of the HOMO.
However in contrast to O$_2$, for a similar range of energy the DCS
of CO$_2$ shows a double-hump structure with the minimum shifting to
a larger angle as energy increases. These features can also be seen
in the phase of the transition dipoles [Fig.~5(d)], where the phase
jump by almost $\pi$ is observed near the minima in the cross
section. The two ``humps'' also behave quite differently. Within the
range of energy from H21 to H39, the small hump at small angles
increases and the big hump at large angles decreases with increasing
energy. We note that the phase jump is smaller than $\pi$ as the
cross section does not go to zero at the minimum between the two
humps. In this paper we will limit ourselves to application of the
QRS to O$_2$ and CO$_2$. Nevertheless it is worthwhile to point out
that the case of N$_2$ is even more complicated than CO$_2$, with
almost no regular behavior found in the energy range presented here,
see Fig.~5(e) and (f). Note that the sharp variation  of cross
section from H17 to H25 is due to the presence of the well-known
shape resonance in N$_2$. The shape resonance is accompanied by a
rapid phase change in the same energy region.

We have seen that these three molecular systems behave totally
differently in photoionization. We show in Sec.~IV that such
differences can be seen from the HHG spectra.

\subsection{Wave packets from molecular targets: example of O$_2$}

In subsection III(C) we have shown evidence to validate the QRS for
atomic targets. For molecular targets some benchmark TDSE results
are available only for H$_2^+$ \cite{lein04,kamta05,telnov07}. In
fact successful application of the QRS and detailed comparisons with
the TDSE results for this system has been reported in \cite{H2+}. In
this subsection for completeness we show here an example of the wave
packets from O$_2$ and Xe, which have nearly identical ionization
potentials (12.03 eV for O$_2$ and 12.13 eV for Xe). In Fig.~6(a) we
compare the HHG spectra from O$_2$ aligned at 10$^{\circ}$ and
70$^{\circ}$,  with that from Xe. These results were obtained from
the SFA, with 1600 nm laser of $1\times 10^{14}$ W/cm$^2$ intensity
and 20 fs duration (FWHM). The long wavelength is used here in order
to compare the wave packets in the extended HHG plateau. Clearly,
the HHG spectra look quite different. However, the wave packets,
extracted by using Eqs.~(\ref{wp-SFA}) and (\ref{wp-ref}), shown in
Fig.~6(b) are almost identical. For clarity we have shifted the
curves vertically in both Figs.~6(a) and (b). The QRS results for
the HHG are shown in Fig.~6(c). The changes in the slope of the QRS
yields as compared to that of the SFA in Fig.~6(a) can be easily
seen, which reflects the differences of the exact and the PWA
photoionization DCS.

\begin{figure}
\mbox{\rotatebox{0}{\myscalebox{
\includegraphics{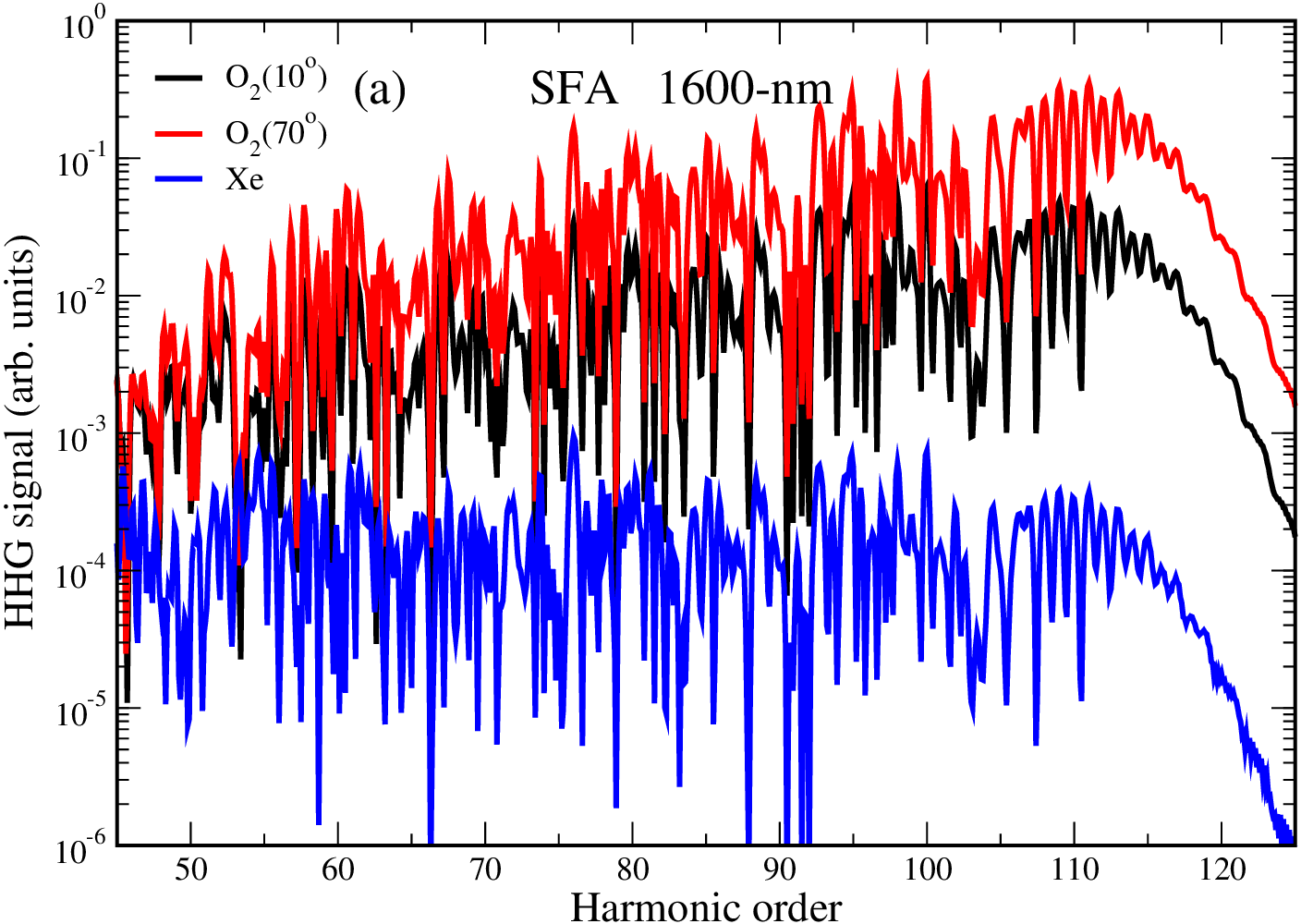}}}}

\mbox{\rotatebox{0}{\myscalebox{
\includegraphics{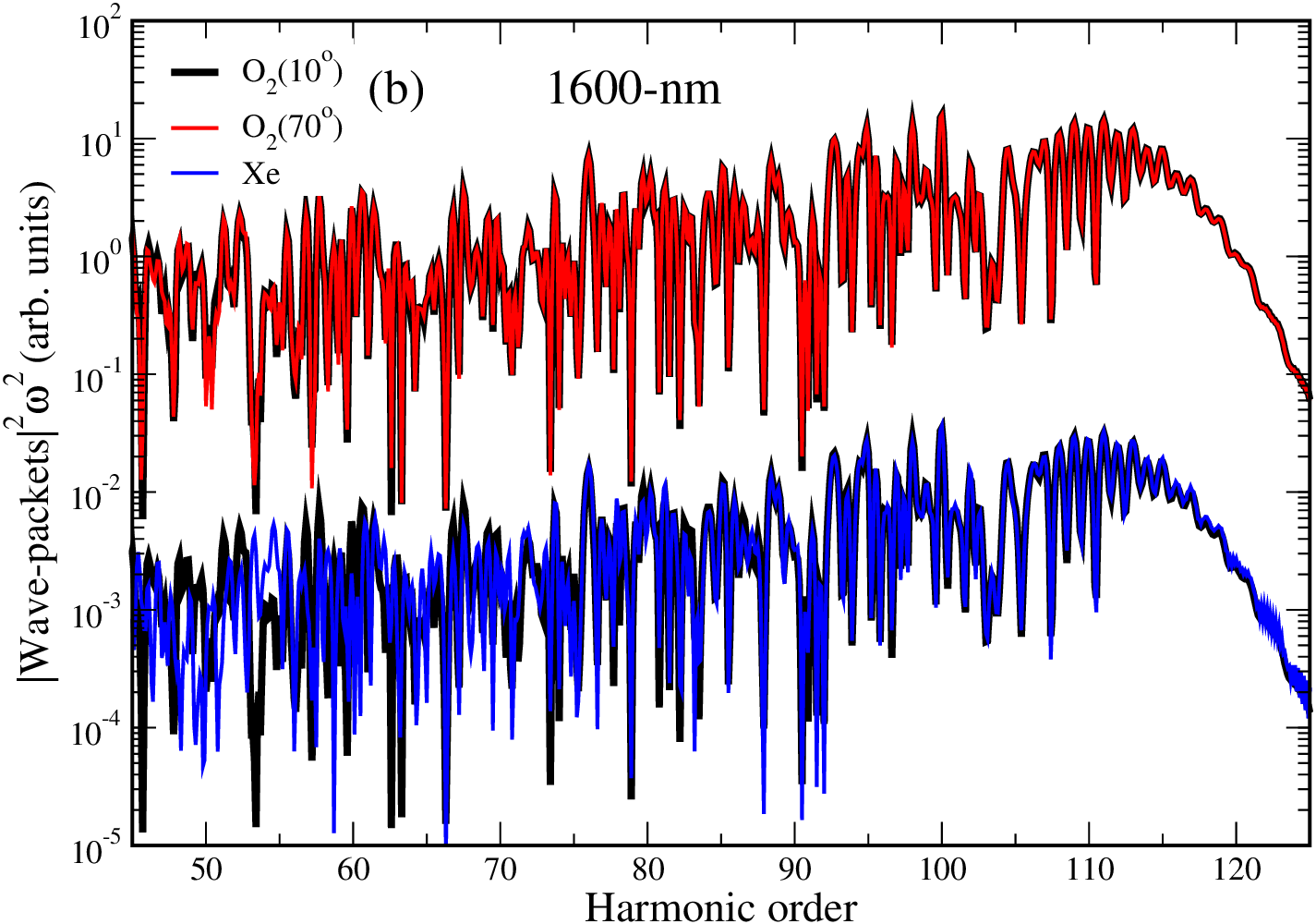}}}}

\mbox{\rotatebox{0}{\myscalebox{
\includegraphics{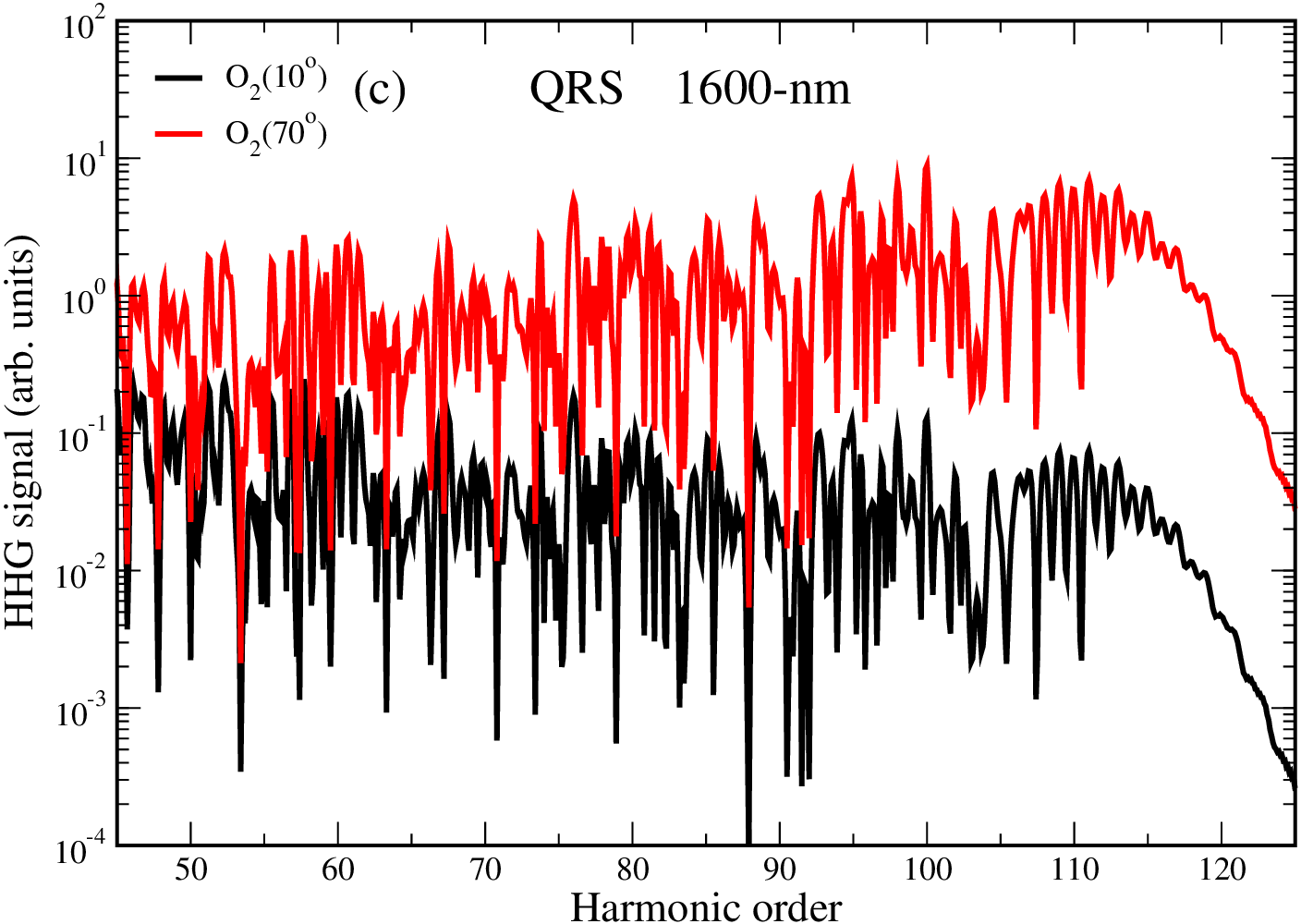}}}}

\caption{(Color online) (a) Comparison of the HHG spectra obtained
from the SFA for O$_2$ aligned at 10$^{\circ}$ and 70$^{\circ}$, and
Xe under laser pulse with wavelength of 1600 nm, intensity of
$1\times 10^{14}$ W/cm$^2$ and duration of 20 fs. For clarity, we
have shifted Xe data vertically. (b) Comparison of the electron
returning wave packets extracted from (a). Wave packet for O$_2$ at
10$^{\circ}$ is renormalized to that of 70$^{\circ}$ and Xe and
plotted twice. (c) The HHG spectra from O$_2$ aligned at
10$^{\circ}$ and 70$^{\circ}$ from the QRS.}
\end{figure}

This example clearly shows that in order to calculate the HHG
spectra for a fixed alignment within the QRS one can use the wave
packet from any other alignment angle or from a reference atom, with
an overall factor accounting for the differences in tunneling
ionization rates [see Eq.~(\ref{normal})]. This feature enables us
to avoid the possible difficulties associated with direct extraction
of the wave packet at the energies and alignments where the
transition dipole in the PWA vanishes. For example, in case of
N$_2$, it is more convenient to use the wave packet extracted for
alignment angle of $90^{\circ}$, as the transition dipole in the PWA
does not vanish. Furthermore, if a reference atom is used, one can
also extract the wave packet from the solution of the TDSE, as has
been done for H$_2^+$ \cite{H2+}.

\section{Comparisons with experiments}

In this section we will compare the QRS results with the recent HHG
measurement from partially aligned molecules. Below we will look at
two examples of molecular O$_2$ (with internuclear distance at
equilibrium $R=2.28$ a.u.) and CO$_2$ ($R=4.38$ a.u. between the two
oxygen centers), which can also be seen as ``elongated'' O$_2$. HHG
from aligned molecules are typically measured experimentally with
the pump-probe scheme. In this scheme, non-adiabatic alignment  of
molecules is achieved by exposing them to a short and relatively
weak laser pulse (the pump) to create a rotational wave packet. This
wave packet rephases after the pulse is over and the molecules are
strongly aligned and anti-aligned periodically at intervals
separated by their fundamental rotational period \cite{seideman}. To
observe the alignment dependence of HHG, a second short laser pulse
(the probe) is then used to produce HHG at different short intervals
when the molecules undergo rapid change in their alignment. A
slightly different setup is done by changing the relative angle
between the pump and probe polarizations, but measure the HHG at
fixed time delays, typically at maximal alignment (or
anti-alignment) near half-revival. Theoretically, induced dipoles
from the QRS need to be calculated first for a fixed molecular
alignment. The results will then be convoluted with the molecular
distribution following the theory presented in Sec.~II(F).

\subsection{HHG from aligned   O$_2$ molecules}

In Sec.~III(D) we showed harmonic spectra generated with 1600-nm
laser. This was used in order to compare the wave packets in an
extended plateau. To compare with available experiments calculations
in this subsection are performed with 800-nm laser. First we show in
Fig.~7(a) the HHG yields for some selected harmonics for {\em fixed}
molecular axis. Calculations are done with the QRS, for a 30-fs
pulse with a peak intensity of $2\times 10^{14}$ W/cm$^2$. The
yields are maximal if molecules are aligned at about $45^{\circ}$
with respect to the probe polarization, and vanish at $0^{\circ}$
and $90^{\circ}$ due to $\pi$-symmetry of the HOMO. This result
resembles closely the behavior of the photoionization DCS's shown in
Fig.~5(a), but with the relative magnitudes changed, reflecting the
influence of the returning wave packets. The yields convoluted with
the partial distribution with maximally aligned ensemble at
half-revival are shown in Fig.~7(b), as functions of relative angle
between pump and probe polarizations. Interestingly, the yields are
now peaked near $0^{\circ}$, that is, when the pump and probe
polarizations are parallel. This result is consistent with the data
by Mairesse {\it et al} \cite{mairesse08} (see their Fig.~2) and
with the earlier measurements by Miyazaki {\it et al}
\cite{miyazaki05}. Note that the yield is quite insensitive to the
alignment angle. However, the contrast can be increased with a
better alignment. Our simulations for alignment are carried out
using the pump laser intensity of $5\times 10^{13}$ W/cm$^2$ and 30
fs duration, with a rotational temperature of 30 K, the same as used
in the experiment. We note that the convoluted yields from the
earlier SFA results by Zhou {\it el al} \cite{zhou05b} show peaks
near $45^{\circ}$ with similar laser parameters. Within the QRS we
found that only with higher degrees of alignment the peaks shift to
about $45^{\circ}$. This can be achieved, for example, with the same
pump laser intensity, but with a longer duration.

\begin{figure}
\mbox{\rotatebox{0}{\myscaleboxd{
\includegraphics{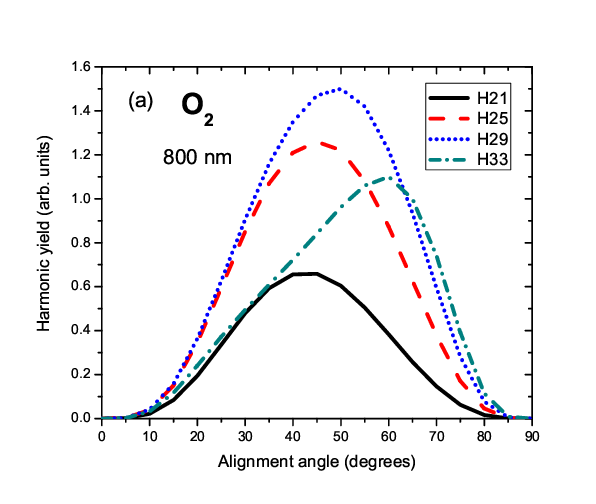}}}}
\mbox{\rotatebox{0}{\myscaleboxd{
\includegraphics{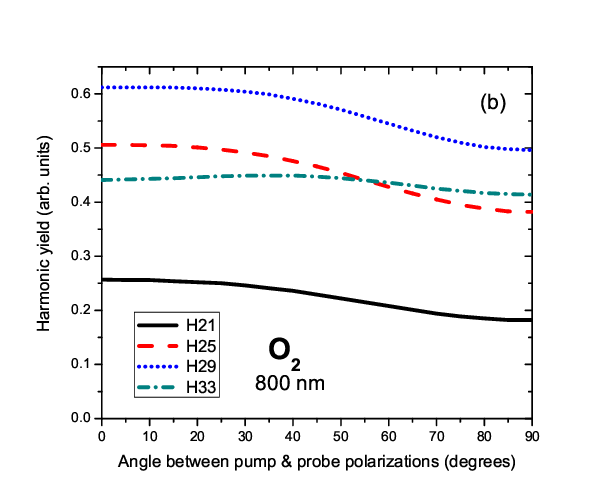}}}}
\caption{(Color online) Alignment dependence of harmonic yields from
O$_2$ harmonics H21, H25,H29, and H33 for fixed molecular axis (a)
and convoluted with molecular alignment distribution (b).}
\end{figure}

Next we compare in Fig.~8 harmonic yields from H17-H35 as a function
of delay time near quarter- and half-revivals from the QRS and the
SFA calculations with the experiments by Itatani {\it et al}
\cite{itatani05}. Calculations were done with laser parameters taken
from Ref.~\cite{itatani05} and the yields have been normalized to
that of the isotropic distribution. Clearly both theories reproduce
quite well the behavior of the experimental curve, which follows
$<\sin^22\theta>$, shown in the upper panel. A closer look reveals
that the QRS result agrees better with the experiment and the SFA
tends to overestimate the yield near maximal alignment and
underestimates it near maximal anti-alignment. In fact, similar
quantitative discrepancies between the SFA and experiments can be
seen in an earlier work by Madsen {\it et al} \cite{madsen06} for
both O$_2$ and N$_2$. Note that the discrepancies between the
simulations and the experiment along the delay time axis could be
due to inaccuracy in the determination of laser intensities and
temperature used in the experiment.
\begin{figure}
\mbox{\rotatebox{0}{\myscaleboxd{
\includegraphics{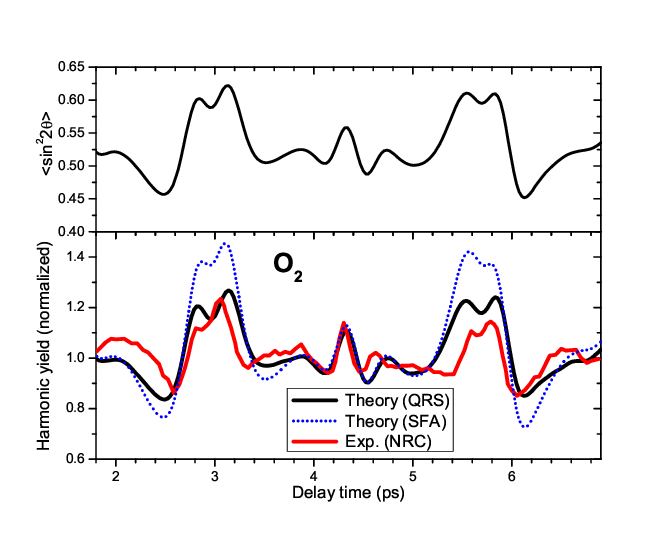}}}}
\caption{(Color online) Comparison of O$_2$ time evolution of
(normalized) harmonic yields H17-H35 from the QRS, SFA and
experiment, near quarter and half-revivals. The experimetal data and
laser parameters are taken from Itatani {\it et al}
\cite{itatani05}. $<\sin^22\theta>$ is shown in the upper panel for reference.}
\end{figure}
We have seen so far only quantitative improvements of QRS over the
SFA for O$_2$. The situation is completely different for CO$_2$,
presented in the next subsection, where the two theories predict
qualitatively different results.

\subsection{HHG from aligned  CO$_2$ molecules}

First we show in Fig.~9(a) theoretical HHG yields for selected
harmonics from H21 up to H39, obtained from the QRS for fixed angles
between the molecular axis and the polarization of the probe pulse.
The yields resemble closely the photoionizations DCS of Fig.~5(c),
with the double-hump structures seen quite clearly for the higher
harmonics above H31. For the lower harmonics, the smaller humps
nearly disappear. To compare with experiment, average over the
molecular alignment needs to be performed. We take the alignment
distribution for a pump-probe delay time at the maximum alignment
near half-revival. The convoluted yields are presented in Fig.~9(b),
as functions of the angle between the pump and probe polarizations.
Clearly, due to the averaging over the molecular alignment
distributions, the angular dependence of HHG is smoother as compared
to fixed alignment data in Fig.~9(a). These results are consistent
with recent experiments \cite{boutu,mairesse08,jila-private}, which
show enhanced yields for large alignment angles and minima near
$30^{\circ}$ for harmonics above H31. We note that our results also
resemble the data for the induced dipole retrieved from mixed gases
experiments by Wagner {\it et al.} \cite{jila07} (see their Fig.~4).
To have a more complete picture of the HHG yields, we show in
Fig.~9(c) a false-color plot of HHG yield as a function of harmonic
orders and the angle between the pump and probe lasers polarization
directions, at the delay time corresponding to the maximum alignment
near the half-revival. Clearly the yield has quite pronounced peak
at large alignment angle. However, the most pronounced feature in
Fig.~9(c) is a minimum at small angles, which goes to larger angles
as with increasing harmonic orders. Within the QRS the origin of
this minimum can be directly traced back as due to the minimum in
the photo-recombination DCS's. Our results are comparable with the
recent measurement by Mairesse {\it et al} \cite{mairesse08} (see
their Fig.~2), which were carried out with a slightly different
laser parameters. Interestingly, our results resemble closely with
the theoretical calculations by Smirnova {\it et al}
\cite{smirnova09}, which include contributions from the HOMO, HOMO-1
and HOMO-2 orbitals [see their Fig.~1(b), plotted in a x-y plot]. In
our calculations, only the HOMO is included.
We note that the position of the minima can be slightly shifted
depending on the actual experimental laser setup. The above
calculations for HHG were done with a 25-fs probe laser pulse with
intensity of $2.5\times 10^{14}$ W/cm$^2$. Alignment distribution
was obtained following the method is Sec.~II(F), with a $120$-fs
pump laser pulse with intensity of $0.55\times 10^{14}$ W/cm$^2$.
Rotational temperature is taken to be $105$ K. These parameters are
taken from the recent experimental setup by Zhou {\it et al}
\cite{jila08}.

\begin{figure}
\mbox{\rotatebox{0}{\myscaleboxc{
\includegraphics{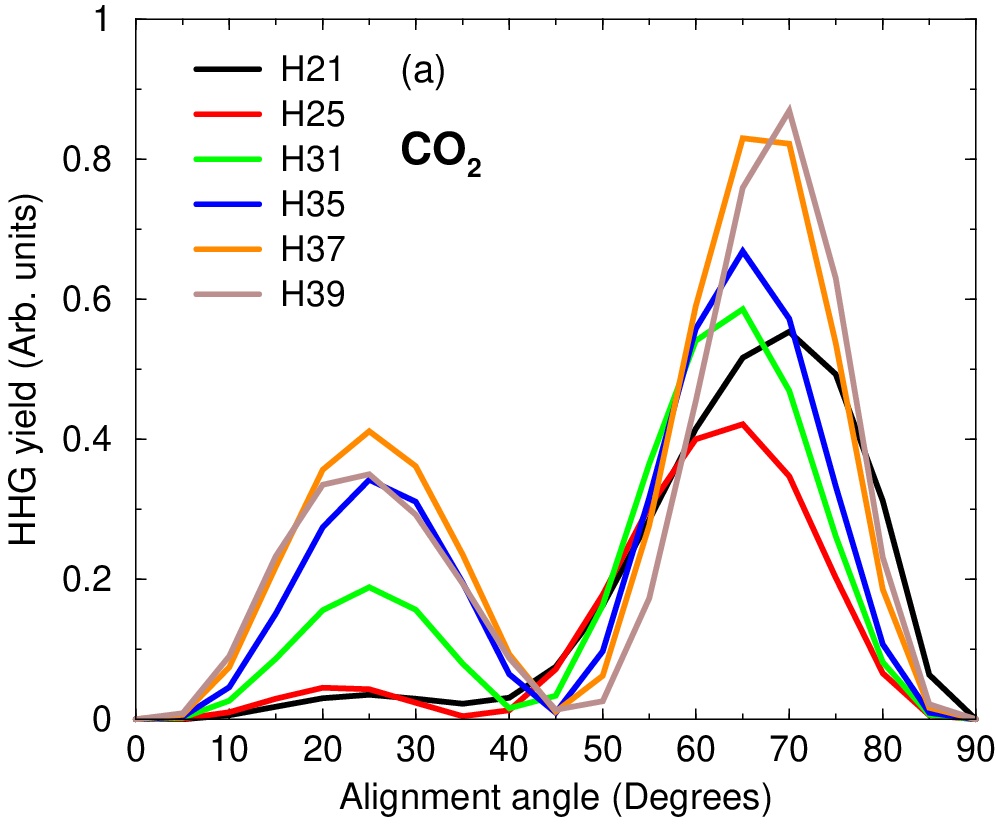}}}}
\mbox{\rotatebox{0}{\myscaleboxc{
\includegraphics{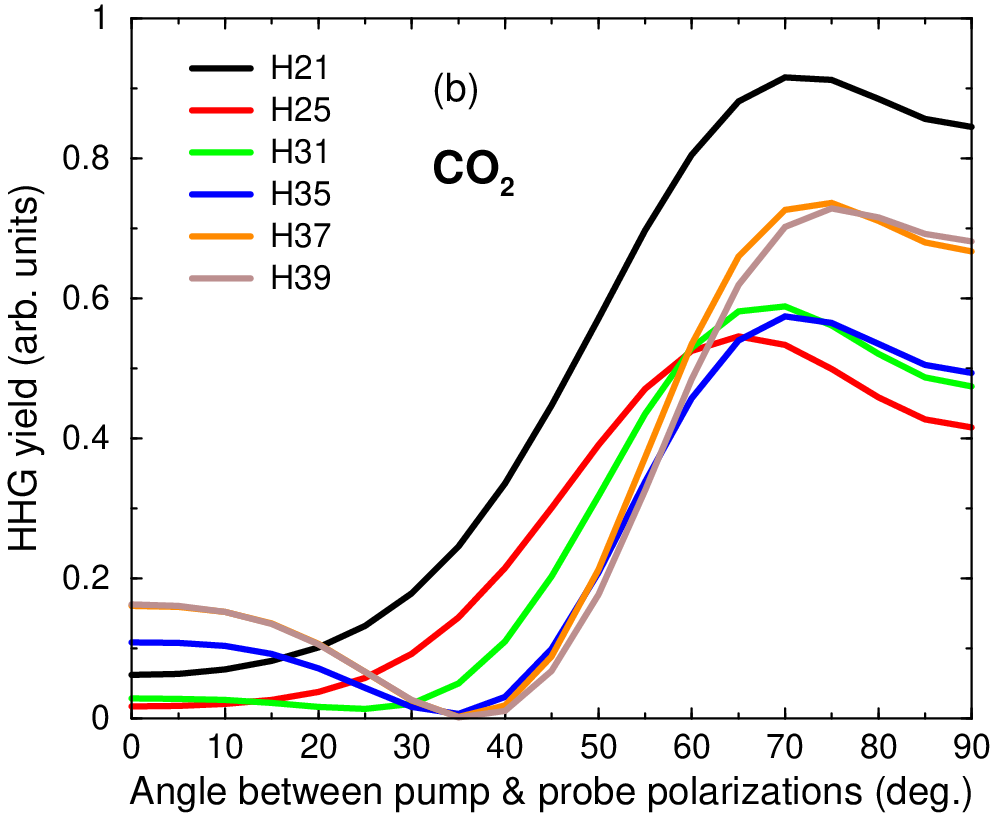}}}}
\mbox{\rotatebox{0}{\myscaleboxc{
\includegraphics{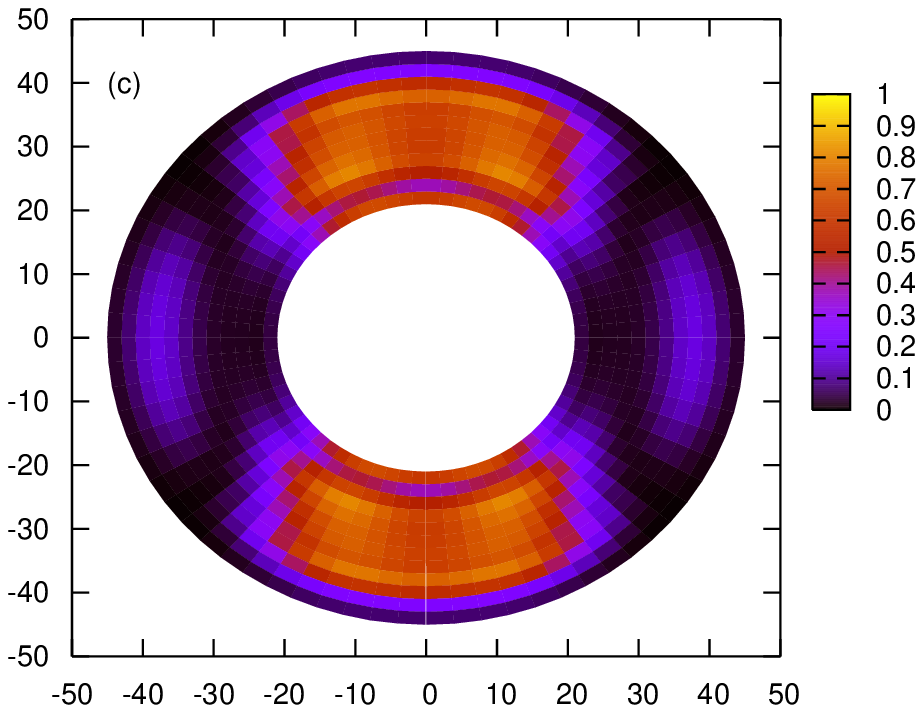}}}}
\caption{(Color online) HHG yield as a function of fixed angle
between molecular axis and laser polarization (a) and as a function
of angle between pump and probe laser polarization directions (b).
False-color plot of HHG yield for a harmonic range between H20 and
H42 (c). In (c) the probe laser is horizontally polarized. Alignment
distribution is chosen at the maximum alignment near half-revival.
The laser intensity and duration are of $0.55\times 10^{14}$
W/cm$^2$, and $120$ fs for the pump pulse, and $2.5\times 10^{14}$
W/cm$^2$ and $25$ fs for the probe pulse. Rotational temperature is
taken to be $105$ K.}
\end{figure}

Comparison of theoretical HHG amplitudes for a few angles between
pump and probe polarizations from $0^{\circ}$ to $25^{\circ}$ is
shown in Fig.~10(a) together with the experimental data in
Fig.~10(b), taken from Mairesse {\it et al} \cite{mairesse08}. The
experimental data have been renormalized to a smoothed experimental
wave packet from Ar under the same laser field. This wave packet is
a decreasing function of harmonic order, therefore the renormalized
signals in Fig.~10(b) are enhanced at higher orders compared to
lower orders. No such renormalization was done for the theoretical
data. Nevertheless, reasonably good agreement with the experimental
data can be seen, including the shift of the minimum position to
higher harmonics as alignment angle increases. The theoretical data
seem to have more structures than the experimental ones. That might
be due to the fact that that we use a single intensity, single
molecule simulation. It is known that macroscopic propagation in
general tends to smooth out the HHG spectra.

\begin{figure}
\mbox{\rotatebox{0}{\myscaleboxd{
\includegraphics{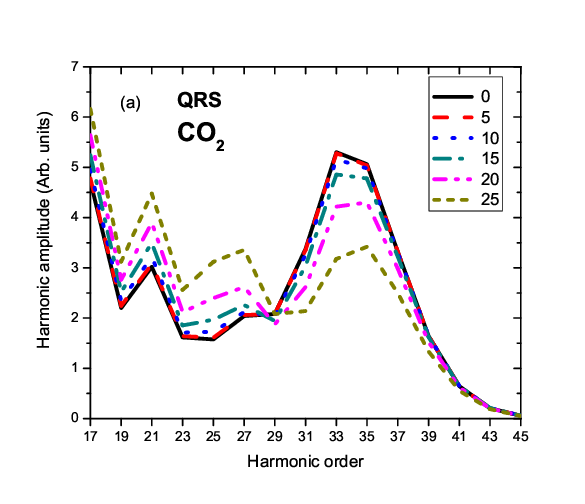}}}}
\mbox{\rotatebox{0}{\myscaleboxd{
\includegraphics{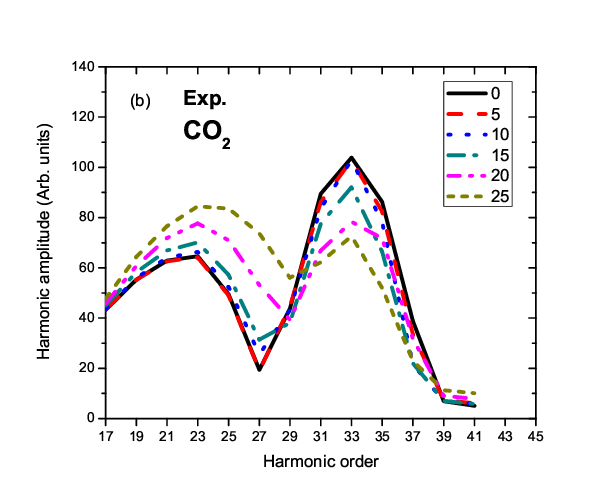}}}}
\caption{(Color online) Harmonic amplitudes for CO$_2$ for a few
angles from $0^{\circ}$ to $25^{\circ}$. The experimental data by
Mairesse {\it et al} \cite{mairesse08} are also shown for comparison
(b). The experimental data have been renormalized to a smoothed
experimental wave packet from Ar under the same laser field. }
\end{figure}

As found in Zhou {\it et al} \cite{jila08}, the most dramatic
feature can be seen near 3/4-revival, where the molecule can be most
strongly aligned. Our QRS results are shown in Fig.~11 for the same
harmonics that have been analyzed in \cite{jila08} (see their
Fig.~2). (We note here that according to the authors of
\cite{jila08}, the harmonic orders should be properly shifted down
by two harmonic orders, as compared to the ones given in their
original paper). Theoretical results were carried out with laser
parameters taken from \cite{jila08}. As can be seen from our results
for the lower harmonics H21 and H25, the yields follow the inverse
of the alignment parameter $\langle\cos^2\theta\rangle$, shown in
Fig.~11(a). However, for the higher harmonics, an additional peak
appears right at the delay time corresponding to the maximum
alignment. The peak starts to appear near H31 and gets more
pronounced with increasing harmonic orders. This behavior is in
quantitatively good agreement with the measurements by Zhou {\it et
al.} \cite{jila08}. In Zhou {\it et al}, the experimental data were
fitted to the two-center interference model, by using a
three-parameter least-square fitting procedure. Within the QRS, no
fitting is needed. Furthermore, one can trace back the origin of the
time-delay behavior based on the two-hump structure of the
photoionization DCS. Indeed, Fig.~5(b) shows that for the lower
harmonics H21 and H25 the cross sections at large angles are much
larger than at small angles so the HHG yield is inverted with
respect to $\langle\cos^2\theta\rangle$, for which small angles
dominate. This fact has been known before \cite{atle06}. For the
higher orders, Fig.~5(b) indicates that the smaller hump at small
angles become increasingly important. Qualitatively, this explains
why the HHG yields for H31 and up show a pronounced peak for the
parallel alignment. In fact, it is even more simpler to explain this
behavior by looking directly at the HHG yields shown in Fig.~9(a).
One can immediately notice that the peak at small angles near
$25^{\circ}$ starts to show up more clearly only near H31. We note
that apart from the additional peaks, which are clearly visible at
3/4-revival, the general behavior of all harmonics from H19 up to
cutoff near H43 shows inverted modulation with respect to
$\langle\cos^2\theta\rangle$. Therefore, in contrast to the
interpretation of Kanai {\it et al} \cite{kanai05},  inverted
modulation is not an unambiguous indication of the interference
minimum, especially for determining the precise position of the
minimum.

\begin{figure}
\mbox{\rotatebox{0}{\myscalebox{
\includegraphics{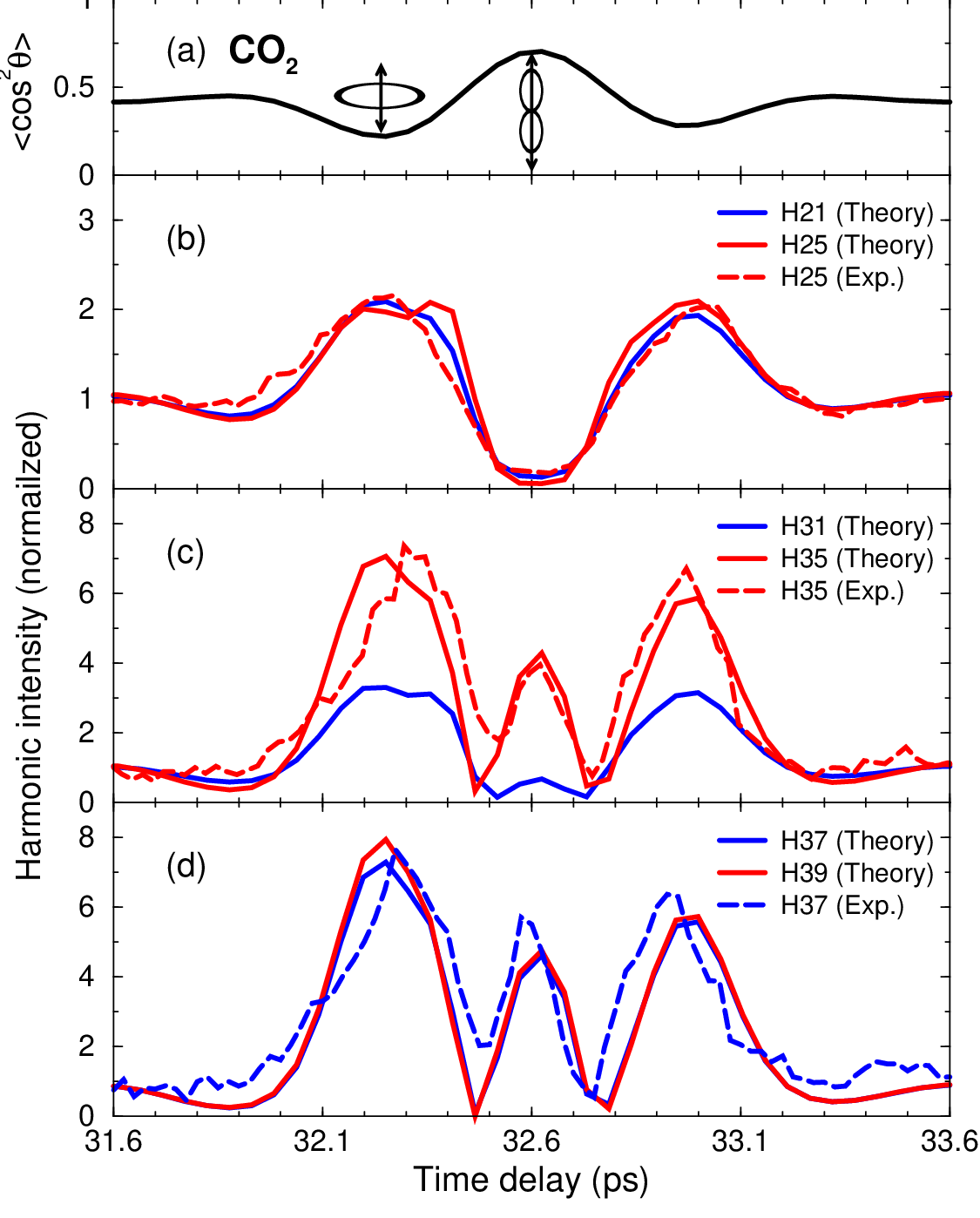}}}}
\caption{(Color online) Normalized  HHG yield (vs isotropically
distributed molecules) from CO$_2$ for selected harmonics as
function of pump-probe delay time near 3/4-revival. The experimental
data are taken from Zhou {\it et al} \cite{jila08}. Theoretical
results were obtained with laser parameters taken from
\cite{jila08}. The alignment parameter $\langle\cos^2\theta\rangle$
is also plotted for reference in the top panel.}
\end{figure}

Since the laser intensity in the experiment is quite high, the
depletion of the ground state needs to be accounted for
\cite{atle06,atle07,xu08}. Therefore discussion about the ionization
rate is in order. Alignment dependence of the ionization rate for
CO$_2$ is still a   subject of debate. In our previous attempts
\cite{atle06,atle07}, the MO-ADK theory \cite{moadk} has been used
to calculate the ionization rate. The MO-ADK theory predicts the
peak in ionization rate near $30^{\circ}$, in agreement with the
results deduced from measured double ionization \cite{KSU-exp}. The
recent experiment by the NRC group \cite{NRC-exp}, however, show
very narrow peak near $45^{\circ}$. Interestingly, the MO-SFA
predicts a peak near $40^{\circ}$. However the MO-SFA theory is
known to underestimate the ionization rate. In order to correct the
MO-SFA rate, we renormalize the MO-SFA rate to that of the MO-ADK
rate at laser intensity of $10^{14}$ W/cm$^2$, which gives a factor
of $10$. Note that the same correction factor has been found for the
SFA ionization from Kr, which has almost the same ionization
potential as for CO$_2$. With the corrected MO-SFA rate, we found
that our simulations give a better quantitative agreement with the
JILA data, than with the MO-ADK rate. On the other hand, calculation
with the uncorrected rate would lead to much more pronounced peaks
right at the maximum alignment near 32.6 ps for H31 and higher
harmonics.

\begin{figure}
\mbox{\rotatebox{0}{\myscaleboxc{
\includegraphics{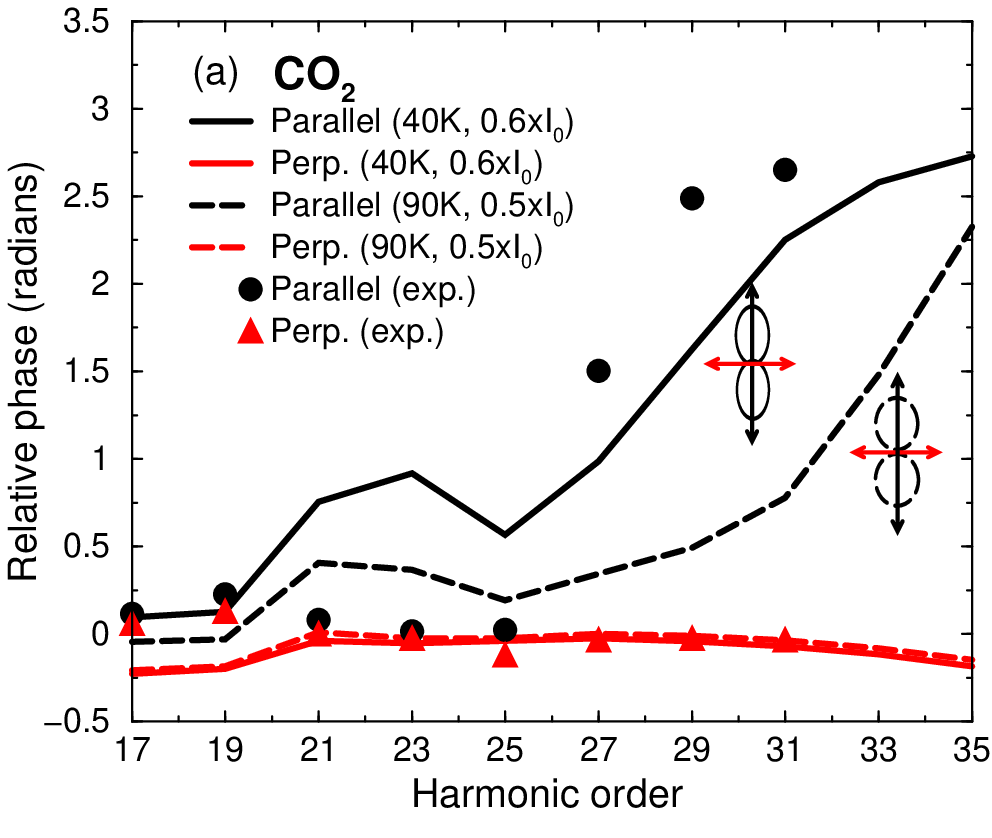}}}}
\mbox{\rotatebox{0}{\myscaleboxc{
\includegraphics{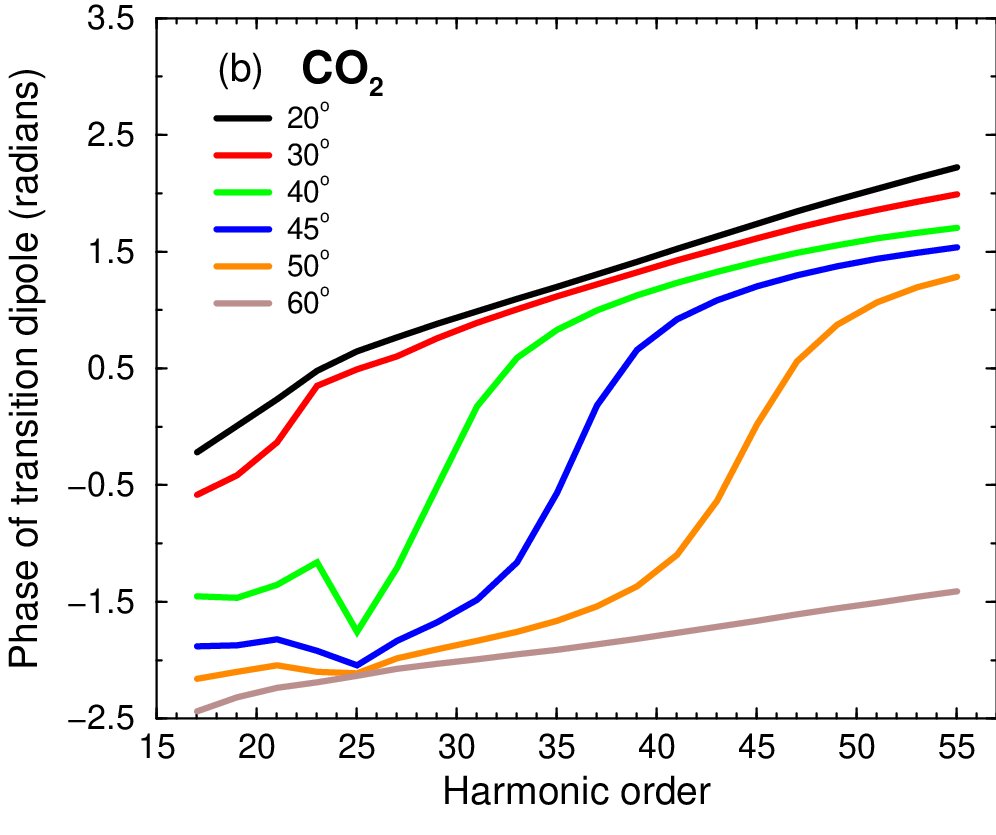}}}}
\caption{(Color online) (a) Harmonic phase (relative to the phase
from Kr) for parallel aligned and perpendicular aligned ensembles,
under two sets of parameters (solid and dashed lines) that lead to
two different degrees of alignment distributions. Experimental data
from Boutu {\it el al.} \cite{boutu} are shown as symbols.
Calculations are carried out with the same probe laser parameters as
in the experiments. (b) Phase of the transition dipole for alignment
angles from $20^{\circ}$ to $60^{\circ}$.}
\end{figure}

We next compare the predictions of the QRS for harmonic phase with
experiments. The phase information can be of great importance. In
particular it has been suggested as  a more accurate way to
determine positions of interference minima in HHG spectra.
Experimentally the harmonics phase can be extracted from
measurements of HHG using mixed gases \cite{boutu,riken08} or
interferometry method \cite{jila08}. In Fig.~12(a),  we show the
recent experimental data of Boutu {\it et al.} \cite{boutu} where
the harmonics phases (relative to that from Kr) are obtained for the
parallel aligned and perpendicularly aligned ensembles, shown by
black and red symbols, respectively. For the latter, the phase does
not change much within H17 to H31. For the parallel aligned
molecules, the phase jump from H17 to H31 was reported to be $2.0\pm
0.6$ radians. Our simulations are shown for two different ensembles,
with alignment distributions confined in a cone angle of
$~25^{\circ}$ (solid lines) and $35^{\circ}$ (dashed lines) at half
maximum. The less aligned ensemble was obtained with the pump laser
parameters and rotational temperature suggested by Boutu {\it et al}
\cite{boutu}. For the perpendicularly aligned molecules, the
relative phase from both ensembles are almost identical and nearly
independent of harmonic orders. This is in good agreement with the
experiment. For the parallel case, our result with the pump laser
parameters suggested in \cite{boutu} shows the phase jump starting
near H31, which only mimics the experimental data. However, the
phase jump slightly shifted to near H27 with the better aligned
ensemble (solid line), bringing the result closer to the experiment.
This indicates that the degree of alignment can play a critical role
in determination of the precise position of the phase jump. A
possible reason for the discrepancy is that the experimental setup
was chosen such that the short trajectory is well phase-matched,
whereas our simulations are carried out at the single molecule level
with contributions from both long and short trajectories. In order
to understand the origin of the phase jump and its dependence on
degree of alignment, it is constructive to analyze transition dipole
phase as a function of harmonic order for fixed alignments. This is
shown in Fig.~12(b) for angular range from $20^{\circ}$ to
$60^{\circ}$. Clearly, the dipole phase shows a phase jump which
shifts to higher order with larger angle. With a better alignment
more contribution comes from small angles and the phase jump shifts
toward smaller harmonic orders. Our calculations are carried out
with a relatively low probe laser intensity of $1.25\times 10^{14}$
W/cm$^2$, the same as used in Boutu {\it et al} experiments
\cite{boutu}, in order to keep ground state depletion insignificant.
As probe intensity increases our results indicate that the phase
jump slightly shifts toward higher harmonic orders. Note that the
phase jump cannot be reproduced within the SFA, where the
insignificant phase difference between CO$_2$ and Kr is caused only
by the small difference in their ionization potentials (13.77 eV vs
14 eV), independently of alignment.

\section{Validity of the Two-center interference model}

Within the QRS theory, the structure of the HHG spectra directly
relates to photo-recombination cross section. In particular, minima
in photo-recombination cross section immediately result in minima in
HHG spectra. It is therefore interesting to compare the positions of
the minima in molecular systems under consideration with the
prediction of the simple two-emitter model by Lein {\it et al}
\cite{lein02}. According to this simple model, the minima satisfy
the relation
\begin{eqnarray}
R\cos\theta &=& (n+1/2)\lambda^{eff}, \quad n=0,1,2, ..., \quad \mbox{symmetric wavefunction}, \\
           &=& n\lambda^{eff},\quad n=1,2, ..., \quad \mbox{anti-symmetric wavefunction}
\end{eqnarray}
where $\lambda^{eff}$  is the ``effective'' wavelength of the
continuum electron defined such that the ``effective'' wave vector
is $k^{eff} = \sqrt{2\Omega}$ , with $\Omega$  being the energy of
the emitted photon. In other words, the energy is shifted by $I_p$
with respect to the usual relation  $k = \sqrt{2(\Omega-I_p)}$.

\begin{figure}
\mbox{\rotatebox{0}{\myscaleboxc{
\includegraphics{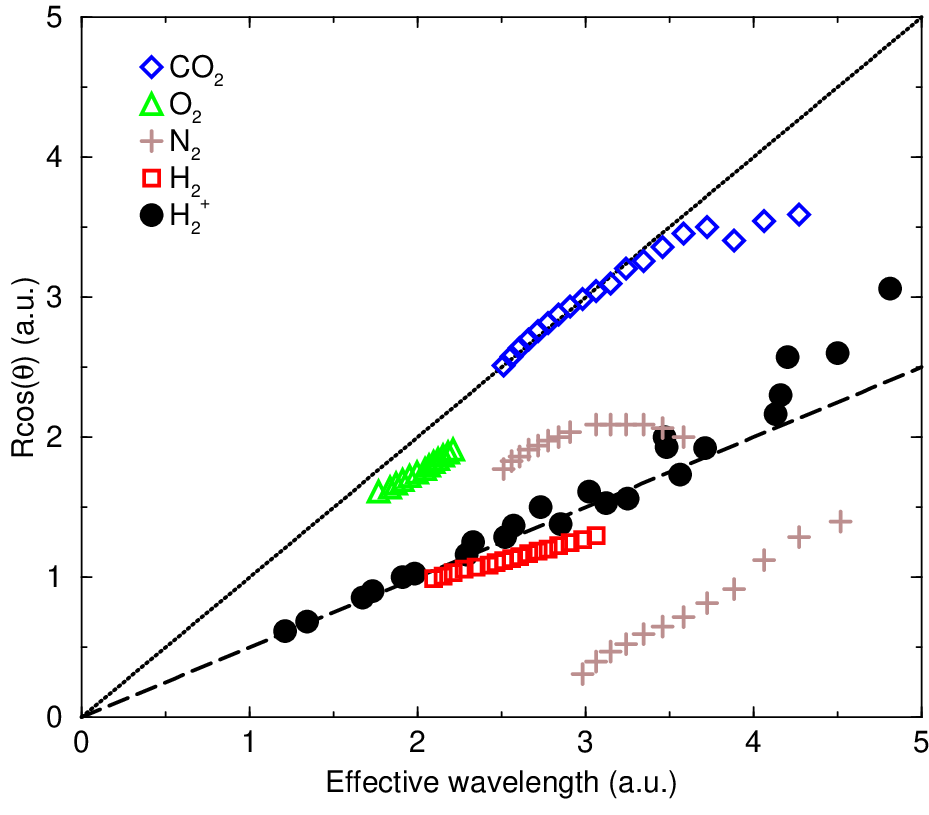}}}}
\caption{(Color online) Projected internuclear distance vs electron
``effective'' wavelength at the minima of the photoionization
differential cross sections for CO$_2$, O$_2$, N$_2$, and H$_2$.
Results for H$_2^+$ with different internuclear separations,
reported in Ref.~\cite{H2+}, are also plotted for comparison.
Predictions of the two emitter model are shown as dashed and dotted
lines for symmetric and anti-symmetric wavefunctions, respectively.}
\end{figure}

In Fig.~13 we show the projected internuclear distance $R\cos\theta$
vs electron ``effective'' wavelength at the minima of the
photoionization differential cross sections for CO$_2$, O$_2$,
N$_2$, and H$_2$. For completeness, we also plot here the results
for H$_2^+$ with different internuclear separations, reported in
Ref.~\cite{H2+}. Remarkably, the minima from CO$_2$ follow the
two-emitter model very well, shown as the dotted line for the case
of anti-symmetric wavefunction (due to the $\pi_g$ symmetry of the
HOMO). This fact has been first observed experimentally in the HHG
measurement from aligned CO$_2$ \cite{kanai05,vozzi05}, and more
conclusive evidences have been shown in
Ref.~\cite{jila08,riken08,boutu}. For O$_2$, which also has the
$\pi_g$ symmetry for the HOMO, the minima start to show up only at
quite high energies (or shorter wavelengths). This is not surprising
since the internuclear distance between the two oxygens is about two
times shorter, as compared to CO$_2$. The case of H$_2$ also agrees
reasonably well with the two-emitter model. However, N$_2$ minima do
not follow any simple pattern. This fact has been noticed earlier by
Zimmermann {\it et al} \cite{zimmermann}. This clearly indicates
that the two-center interference model is not guarantied to work
{\it a priori}.

\begin{figure}
\mbox{\rotatebox{0}{\myscaleboxd{
\includegraphics{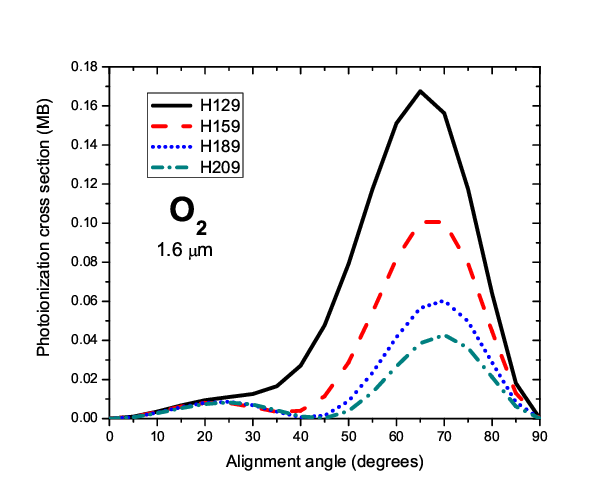}}}}
\caption{(Color online) Diffrential photoionization cross sections
for O$_2$ as functions of the alignment angle. The photon energies
are expressed in units of harmonic orders for 1600-nm laser.  }
\end{figure}

Finally we show that the minima in the photoionization cross section
of O$_2$ can be observed in HHG experiments. Since the minima occur
at quite high energies, it is better to use a driving laser with
long wavelength of 1600 nm \cite{midorikawa-prl08}, as the cutoff
will be extended to much higher energies without the need of using
high laser intensities. In Fig.~14 we plot the photoionization cross
sections for harmonic orders  of 129, 159, 189 and 209, as functions
of alignment angle. Note that the lower orders are already given in
Fig.~5 (but in units of harmonics for 800 nm). The cross sections
show a clear minimum for H159 near $37^{\circ}$, which slowly moves
to larger angles for higher harmonics. This picture resembles the
behavior for the CO$_2$ case, but at much higher energies. Similar
to CO$_2$, a pump-probe scheme can be used to investigate the time
delay behavior of the HHG yield.

\begin{figure}
\mbox{\rotatebox{0}{\myscaleboxd{
\includegraphics{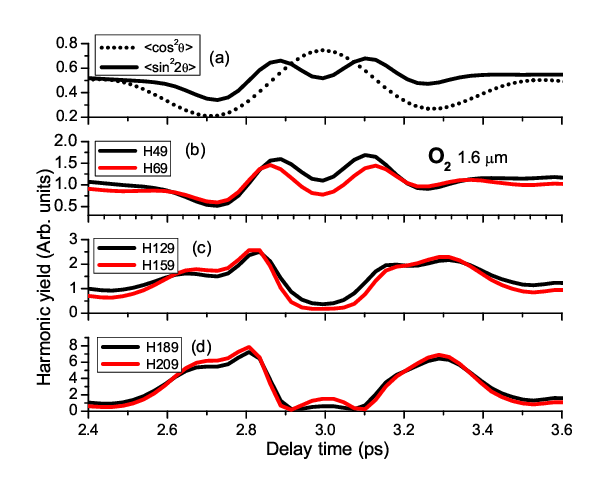}}}}
\caption{(Color online) HHG yield  from O$_2$ for different
harmonics as function of pump-probe delay time near quarter-revival.
$\langle\sin^2(2\theta)\rangle$ and $\langle\cos^2(\theta)\rangle$
are also plotted for reference (a). Calculations are performed with
a 1600-nm laser pulse with peak intensity of $2\times 10^{14}$
W/cm$^2$ and 20 fs duration.}
\end{figure}

In Fig.~15 we show a typical behavior of the HHG yield as function
of delay time between pump and probe laser pulses for H49 and H69 in
the lower plateau, H129 and H159 in the middle plateau, and H189 and
H209 near the cutoff. The calculations were carried out with a
1600-nm laser pulse with peak intensity of $2\times 10^{14}$
W/cm$^2$ and 20 fs duration (FWHM). For the pump we use 800-nm pulse
with $4\times 10^{13}$ W/cm$^2$ and 120 fs duration. We only focus
on the results near the quarter-revival (near 3 ps), when the
molecule can be most strongly aligned. This is in contrast to CO$_2$
where most strong alignment is achieved near 3/4-revival. The
difference is caused by the different symmetries of the total
electronic ground state wavefunctions, as discussed in Sec.~II(F).
Clearly the behaviors are quite different for these energy ranges.
In the lower plateau, the HHG yield follows
$\langle\sin^2(2\theta)\rangle$. This behavior is the same as for
the case of 800-nm laser. In the middle plateau, the behavior starts
to look similar to that of CO$_2$, which follows the inverted of
$\langle\cos^2(\theta)\rangle$. This reflects the fact that the
photoionization DCS peaks have shifted to larger angles for higher
photon energy. Finally, for even higher harmonics an additional peak
occurs right at the quarter-revival. This is also similar to the
case of CO$_2$ near H31 for 800-nm, where the transition dipole goes
through a minimum. Note that a quite high degree of alignment is
needed in order to observe these additional peaks, which is not as
pronounced as in CO$_2$ near 3/4-revival.

\section{Summary and Outlook}

In this paper we give detailed description of the quantitative
rescattering theory (QRS) applied for high-order harmonic generation
(HHG) by intense laser pulses. Although HHG has been described by
the three-step model \cite{corkum} in its many versions, including
the Lewenstein model \cite{lewenstein}, quantum orbits theory
\cite{milosevic-review} since the mid 1990's, the calculated HHG
spectra are known to be inaccurate, especially in the lower plateau
region, as compared to  accurate results from the TDSE. At the same
time, numerical solution of the TDSE for molecules in intense laser
pulses still remains a formidable challenge and   has been carried
out only for the simplest molecular system H$_2^+$ so far. The QRS
has been shown to provide a simple method for calculating accurate
HHG spectra generated by atoms and molecules.   The essence of the
QRS is contained in Eq.~(\ref{QRS-model-1}) which states that
harmonic dipole can be presented as a product of the returning
electron wave packet and exact photo-recombination dipole for
transition from a {\em laser-free} continuum state back to the
initial bound state. The validity of the QRS has been carefully
examined by checking against accurate results for both harmonic
magnitude and phase from the solution of the TDSE for atomic targets
within the single active electron approximation. For molecular
targets, the QRS has been mostly tested for self-consistency in this
paper. More careful tests have been reported earlier in
Ref.~\cite{H2+} for H$_2^+$, for which accurate TDSE results are
available. The results from the QRS for molecules are in very good
agreement with available experimental data, whereas the SFA results
are only qualitative at best.

Here are a number of the most important results of the QRS.

(i) The wave packet is largely independent of the target and
therefore can be obtained from a reference atom, for which numerical
solution of the TDSE can be carried out, if needed. It can also be
calculated from the SFA. The shape of the wave packet depends on the
laser parameters only, but its magnitude also depends on the target
through the ionization probability for electron emission along laser
polarization direction.

(ii) By using the factorization of HHG and the independence of the
wave packet of the target structure, accurate photo-recombination
transition dipole and phase can be retrieved from  HHG experiments.
This has been demonstrated for atomic targets, both theoretically
\cite{toru,atle08} and experimentally \cite{minemoto08}, and more
recently for CO$_2$ \cite{atle09}.

(iii) Since the QRS is  almost as simple as the SFA, it can be
useful for realistic simulations of experiments, where macroscopic
propagation needs to be carried out. For such simulations
contributions from hundreds of intensities are needed. Existing
macroscopic progagation simulations have rarely been done beyond the
SFA model even for atomic targets. In the QRS the most
time-consuming part is the calculation of the photoionizaion
transition dipole. However, this needs to be done only once for each
system, independently of laser parameters. The use of the QRS would
significantly improve the SFA macroscopic results, but with the same
computational effort as for the SFA.

(iv) Because of the inherent factorization, the QRS allows one to
improve separately the quality of the wave packet and of the
transition dipole. In particular, one can include many-electron
effect in the photoionization process, which has been performed
routinely in the atomic and molecular photoionization research.

The are several limitations of the QRS. Clearly, the method is not
expected to work well in the multiphoton regime since QRS is based
on the rescattering physics. Wave packets from the SFA becomes less
accurate for lower plateau. We have seen that this has affected both
harmonic yield and phase. This is probably due to the fact that the
electron-core interaction has been neglected in the SFA during the
electron propagation in the continuum. Remedy for this effect has
been suggested by including some correction to the semi-classical
action \cite{ivanov96}. We emphasize that the QRS so far only
improves the last step of the three-step model by using exact
photo-recombination transition dipole. Lastly we mention that
similarity of the wave packets from different targets only holds for
low to moderate intensities. Near saturation intensities, when
depletion effect is large, the wave packets from targets with
different ionization rates would be very different. Nevertheless,
one can still use the wave packet from the SFA, provided the
depletion is properly included.

The evidence presented in this work strongly supports that the QRS
provides as a powerful method to obtain accurate   HHG yield and
phase from molecules under intense infrared laser pulses. As the
technology of HHG generation improves, one can count on the
retrieval of accurate photoionization cross section and phase   from
such measurement from aligned molecules. By using a pump beam to
initiate a chemical reaction and a probe laser beam to generate
high-order harmonics in a standard pump-probe scheme, the QRS would
allow the retrieval of the transition dipole magnitude and phase
over a broad range of photon energies, thus paving the way for
extracting the structure information of the transient molecule, to
achieve ultrafast chemical imaging with femtosecond temporal
resolutions. In the meanwhile, with accurate single atom or single
molecule induced dipole moments readily calculated, it is also
possible to examine the effect of macroscopic propagation of the
harmonics in the gaseous medium.

\section*{Acknowledgements}
We thank X. Zhou, N. Wagner, M. Murnane, H. Kapteyn, P. Salieres,
and D. Villeneuve for communicating their results to us and the
valuable discussions. This work was supported in part by the
Chemical Sciences, Geosciences and Biosciences Division, Office of
Basic Energy Sciences, Office of Science, U. S. Department of
Energy. TM was also financially supported by Grant-in-Aid from the
Japan Society for the Promotion Science (JSPS) and the PRESTO
program of the Japan Science and Technology Agency (JST).

\end{document}